\documentclass[%
 reprint,
superscriptaddress,
nofootinbib,
 amsmath,amssymb,
 aps,
floatfix,
]{revtex4-2}

\usepackage{graphicx}
\usepackage{dcolumn}
\usepackage{bm}
\usepackage[colorlinks,allcolors=black]{hyperref}
\usepackage[capitalize]{cleveref}
\usepackage{aas_macros}
\usepackage{amsmath}	
\usepackage{tikz}

\newcommand{\cii}{{[}C\,\textsc{ii}{]}}
\newcommand{\avg}[1]{\left\langle#1\right\rangle}
\DeclareUnicodeCharacter{2032}{$^\prime$}

\usepackage{xcolor}
\usepackage{xspace}

\newcommand{\COBE}{{\it COBE}\xspace}

\newcommand{\PIXIE}{{\it PIXIE}\xspace}
\newcommand{\SPIXIE}{{\it SuperPIXIE}\xspace}

\newcommand{\changeJ}[1]{{#1}}
\newcommand{\changeJfirstround}[1]{{#1}}

\newif\iftrack
\iftrack
\newcommand{\added}[1]{{\bf #1}}
\newcommand{\deleted}[1]{}
\newcommand{\replaced}[2]{{\bf #2}}
\else
\newcommand{\added}[1]{{#1}}
\newcommand{\deleted}[1]{}
\newcommand{\replaced}[2]{{#2}}
\fi
\newif\iftracktwo
\iftracktwo
\newcommand{\addedtwo}[1]{{\bf #1}}
\newcommand{\deletedtwo}[1]{}

\else
\newcommand{\addedtwo}[1]{{#1}}
\newcommand{\deletedtwo}[1]{}

\fi

\begin{document}

\title{Carbon monoxide and ionized carbon line emission global signals: \texorpdfstring{\\}{}foregrounds and targets for absolute microwave spectrometry}

\author{Dongwoo T.~Chung}
 \email{dongwooc@cita.utoronto.ca}
\affiliation{%
 Canadian Institute for Theoretical Astrophysics, University of Toronto, 60 St.~George Street, Toronto, ON M5S 3H8, Canada\\}
\affiliation{Dunlap Institute for Astronomy and Astrophysics, University of Toronto, 50 St.~George Street, Toronto, ON M5S 3H4, Canada
}
\affiliation{\addedtwo{Department of Astronomy, Cornell University, Ithaca, NY 14853, USA}
}
\author{Jens Chluba}
\affiliation{Jodrell Bank Centre for Astrophysics, School of Physics and Astronomy, The University of Manchester, Manchester M13 9PL, U.K.}
\author{Patrick C.~Breysse}
\affiliation{Center for Cosmology and Particle Physics, Department of Physics, New York University, 726 Broadway, New York, NY, 10003, U.S.A.}

\date{\today}

\begin{abstract}
We consider the potential of future microwave spectrometers \changeJfirstround{akin to the Primordial Inflation Explorer (\PIXIE)} in light of the sky-averaged global signal expected from the total intensity of extragalactic carbon monoxide (CO) and ionized carbon (\cii{}) line emission. We start from models originally developed for forecasts of line-intensity mapping (LIM) observations targeting the same line emission at specific redshifts, extrapolating them across all of cosmic time. We then calculate Fisher forecasts for uncertainties on parameters describing relic spectral deviations, the CO/\cii{} global signal, and a range of other Galactic and extragalactic foregrounds considered in previous work. We find that the measurement of the CO/\cii{} global signal with \changeJfirstround{a future CMB spectrometer} presents an \changeJfirstround{exciting} opportunity to constrain the evolution of metallicity and molecular gas in galaxies across cosmic time. From \PIXIE to \changeJfirstround{its enhanced version, \SPIXIE}, microwave spectrometers would have the fundamental sensitivity to constrain the redshift evolution of \changeJ{the} average kinetic temperature and cosmic molecular gas density at a level of 10\% to 1\%, \changeJfirstround{respectively. Taking a spectral distortion-centric perspective,} when combined with other foregrounds, sky-averaged CO/\cii{} emission can mimic $\mu$- and to a lesser extent $y$-type distortions. Under fiducial parameters, marginalising over the CO/\cii{} model parameters increases the error on $\mu$ by \replaced{\changeJfirstround{$\simeq 50\%$}}{$\simeq 86\%$}, and the error on $y$ by \changeJfirstround{$\simeq 10\%$}. Incorporating information from \changeJfirstround{planned} CO LIM surveys can recover some of this loss in precision. Future work should deploy a more general treatment of the microwave sky to quantify in more detail the potential synergies between \PIXIE-like and CO LIM experiments, which complement each other strongly in breadth versus depth, and ways to optimise both spectrometer and LIM surveys to improve foreground cleaning and maximise \changeJfirstround{the science return for each}.
\end{abstract}

\maketitle




\section{Introduction}
\label{sec:intro}
The sky-averaged frequency spectrum of the cosmic microwave background (CMB), as Fixsen and Mather~\cite{FIRASFinal} note, is `sometimes stated' to be `the most perfect blackbody spectrum ever measured', with the Cosmic Background Explorer (\COBE) satellite's Far Infrared Absolute Spectrophotometer (FIRAS) measuring a blackbody temperature of \changeJfirstround{$2.725\pm0.001$\,K \citep{Fixsen1996, Fixsen2009}}. Deviations from the spectrum were 50 ppm (rms) of peak brightness, within measurement uncertainties.

However, we absolutely expect spectral distortions in the CMB monopole \changeJfirstround{\citep[e.g.,][]{Sunyaev2009, Sunyaev2013, Chluba16}, which has driven significant activity in connection with \replaced{the planning of the future space missions \citep{Chluba2021ExA, Chluba2021Voyage}}{planning of future space missions \citep{Chluba2021ExA}}}. As the baryon-photon fluid cools through expansion, rates for scattering \changeJfirstround{and photon emission} processes fall below the rate required to sustain thermalisation against energy infusion. The resulting relic spectral distortions fall into two types, according to most conventions (as in, e.g.,~\citep{Silk1992} and~\citep{Fixsen1996}). The $\mu$-type distortion takes its name from the chemical potential parameter $\mu$ that characterises deviation of the Bose--Einstein photon distribution away from a blackbody function \changeJfirstround{\citep{Sunyaev1970mu, Burigana1991, Hu1993, Chluba2014}}. This occurs as Coulomb and double Compton scattering processes become too inefficient to maintain thermal equilibrium, giving way to normal Compton scattering, which diffuses the photon distribution in frequency space. A $y$-type distortion results even later in time, following the same physics as the thermal Sunyaev--Zel'dovich (tSZ) effect commonly observed through hot clusters, but on a primordial, universal scale \changeJfirstround{\citep{Zeldovich1969}}. We characterise this effect using the Compton $y$ parameter, an integral of the temperature deficit over the Compton scattering-dominated electron optical depth.


The \COBE mission set a remarkable benchmark for measuring such distortions, finding upper limits (95\% CL) of \changeJfirstround{$|y|<1.5\times10^{-5}$} and $|\mu|<9\times10^{-5}$~\citep{Fixsen1996,FIRASFinal}, \changeJfirstround{implying that no more than $\delta \rho/\rho \simeq 6\times 10^{-5}$ of energy could have been released into the CMB}. A subsequent re-analysis of the \COBE data by~\cite{BianchiniFabbian22} deployed improved foreground cleaning techniques to refine the FIRAS limits to $|\mu|<4.7\times10^{-5}$. While the predominant science target for cosmic microwave observations has since been anisotropies with only relative calibration, some Earth-bound experiments like TRIS~\citep{TRIS1,TRIS2} and ARCADE2~\citep{ARCADE2} have followed in \changeJfirstround{FIRAS'} path of absolute microwave spectrometry. Despite being a ground-based experiment, TRIS yielded constraints competitive with the FIRAS limits, for instance reporting $|\mu|<6\times10^{-5}$ (95\% CL). ARCADE2, meanwhile, identified an intriguing excess isotropic radio background whose origin remains very poorly understood~\citep{RSB2}. \changeJfirstround{Most recently, this excess was interpreted as a signal from cosmic string decays, potentially also affecting the interpretation of 21cm observations \citep{Cyr2023CS}.} These examples indicate both the potential for future measurements of $y$ and $\mu$ to move forward beyond current limits, and the discovery space that such absolute spectrometry can access.

Extragalactic emission in the carbon monoxide (CO) rotational transitions and ionized carbon \cii{} fine-structure line finds itself redshifted across the microwave spectrum, serving both as a potential obstacle to improved $y$ and $\mu$ constraints, and as a significant region of the discovery space yet unaccessed. First studied as a foreground for CMB anisotropy measurements~\citep{Righi08}, a number of works have also considered the global signal associated with extragalactic CO and/or \cii{} and compared this spectrum to the amplitude of relic CMB spectral distortions~\citep{Mashian16,Serra16}. In turn, Abitbol \emph{et al.}~\cite{Abitbol17} used the predicted extragalactic CO spectrum from Mashian, Loeb, and Sternberg~\cite{Mashian16} as a template alongside a range of other Galactic and extragalactic foregrounds, showing that they did not prevent future spectrometers from making significant detections of relic spectral distortions.

However, a community emerging around the technique of line-intensity mapping (LIM; see~\citep{LIM2019} and~\citep{BernalKovetz22} for reviews) is also putting significant effort into modelling extragalactic CO and \cii{}, except as the primary target for a range of LIM experiments rather than merely as foreground emission. Over the course of theoretical and experimental work in the past decade~\citep{Lidz11,Gong12,Crites14,Silva15,Yue15,Li16,Keating16,Padmanabhan18,Padmanabhan19,Dumitru19,Sun19,mmIME-ACA,CCATp,SW21a,SW21b,Sun21,Yang22,Cleary22,Padmanabhan22,Bethermin22,CONCERTO,CII_MHIZ}, the LIM community has continued to push forward on empirical modelling motivated by steadily improving information on high-redshift galaxies from ALMA surveys \changeJfirstround{\citep[e.g.,][]{Carilli2016, Decarli20,ALPINELF,Heintz22}} as well as insights from simulations~\citep[e.g.,][]{Lagache_2018,liang2023}. Despite such progress, theoretical work has so far not strongly converged on expectations for the brightness of CO and \cii{} line emission at high redshift, with predictions varying by at least an order of magnitude. Such uncertainty derives from key unknowns in the cosmic history of star formation and interstellar gas, which future LIM observations will uncover. This uncertainty also suggests exciting science prospects for other observations at microwave frequencies, including spectral distortion experiments (see, e.g.,~\citep{Anderson22}).

A path forward exists for absolute microwave spectrometry observations beyond FIRAS, TRIS, and ARCADE2, in the near future through experiments like TMS~\citep{TMS}, BISOU~\citep{BISOU}, and COSMO~\citep{COSMO}, and further forward in time through a potential return to space-based spectrophotometry \changeJfirstround{\citep{PIXIE,Chluba2021Voyage, V2050}}. Such observations will provide complementary information to CO/\cii{} LIM experiments that probe line-intensity anisotropies over more defined volumes at specific redshift ranges. At the same time, as previous works have shown and as the current work will illustrate, the CO/\cii{} global signal imprints itself on the microwave spectrum at a similar level to relic spectral distortions and will be an important foreground to anticipate. As such, rather than stop at predicting the CO/\cii{} global signal or incorporating a fixed CO template as a nuisance to marginalise away, this work aims to both more finely quantify the impact of modelling the CO/\cii{} global signal on spectral distortion constraints, and quantify the fundamental sensitivity of future observations to information about the cosmic history of carbon in star-forming galaxies.

The purpose of the present work is thus twofold, in answering the following questions:
\begin{itemize}
    \item How can spectral distortion surveys constrain the modelling of cosmological CO and \cii{} emission?
    \item How does the modelling of extragalactic CO and \cii{} in turn affect constraints on $y$- and $\mu$-type distortions?
\end{itemize}

We structure the paper as follows. \autoref{sec:methods} outlines our forecasting methodology, focussing primarily on the extragalactic CO and \cii{} models used. \autoref{sec:results} presents the results of this forecasting, with the CO/\cii{} global signal treated as both spectral distortion foreground and science target. Finally,~\autoref{sec:conclus} recaps the key outcomes of the present work and outlines further work needed to realise the potential of future microwave spectrometers.

Unless otherwise stated, we assume base-10 logarithms, and a $\Lambda$CDM cosmology with parameters $\Omega_m = 0.307$, $\Omega_\Lambda = 0.693$, $\Omega_b =0.0486$, $H_0=100h$\,km\,s$^{-1}$\,Mpc$^{-1}$ with $h=0.677$, $\sigma_8 =0.816$, and $n_s =0.967$, consistent with the findings of the \emph{Planck} satellite (based on~\citep{Planck15} but still consistent even with~\citep{Planck18_}). Distances carry an implicit $h^{-1}$ dependence throughout, which propagates through masses (all based on virial halo masses, proportional to $h^{-1}$) and volume densities ($\propto h^3$).

\section{Methods}
\label{sec:methods}
For the most part, this paper reuses the models and methods of~\cite{Abitbol17}. The analysis accounts for deviations from the current known value of $T_\text{CMB}=2.7255$ K,  cumulative $y$-type Sunyaev--Zel'dovich distortions including relativistic corrections, primordial $\mu$-type distortions, thermal dust, the cosmic infrared background (CIB; extragalactic dust), Galactic synchrotron emission, Galactic free-free emission, extragalactic CO emission, and anomalous microwave emission (AME). We refer the reader to that paper and references therein for details of these distortions and foregrounds outside of extragalactic carbonic lines.

The primary changes we make are alterations to the last two components and the addition of extragalactic \cii{} emission. The changes to the AME implementation are minor; we simply re-derive a template from the two-component spinning dust model used by~\cite{Planck15_X}, calculating the spectrum in each pixel and averaging across 70\% of the sky (masking out the 30\% brightest pixels in AME at 30 GHz). We continue to allow only one free parameter, an overall dimensionless scaling $A_\text{AME}$ of the resulting template spectrum.

The changes to the extragalactic line emission are greater. We expand significantly on the approach of~\cite{Abitbol17}, which used the CO model of~\cite{Mashian16} as a fixed template with some overall scaling parameter, by deploying models motivated by the molecular gas mass content of constituent molecular clouds and the neutral gas mass of galaxies. The model bears similarities to the physically motivated approach of the LIMFAST code~\citep{LIMFAST1,LIMFAST2}, but is a decidedly empirical halo model of line emission while injecting certain physical elements for a self-consistent treatment of the CO rotational ladder.

For the remainder of the section, we first detail the CO and \cii{} line emission models used in this work in~\autoref{sec:COmodel} and~\autoref{sec:CIImodel}, extrapolations of models previously developed in LIM contexts for line emission from specific redshifts. We then explain in~\autoref{sec:fishermethods} the Fisher forecasts used to evaluate these line emission components against future experimental sensitivities.
\subsection{Carbon monoxide line emission model}
\label{sec:COmodel}
To model CO emission from a wide range of redshifts, we consider an extrapolation of Model B of~\cite{COMAPTDX}. This model uses the off-the-shelf Radex non-LTE radiative transfer code~\citep{Radex} with a simple prescription for molecular cloud occupation of halos to consider a halo model of CO line emission covering multiple rotational transitions.

We assume that the average molecular cloud has molecular mass $2.2\times10^4\,M_\odot$, radius $r_\text{MC}=8.6$ pc and H$_2$ number density $n_\text{H2}=110$\,cm$^{-3}$. We also assume that the kinetic temperature $T_k$ roughly follows the dust temperature as a function of redshift, and for this use the fitting result of~\cite{Viero22}:
\begin{equation}
    \frac{T(z)}{\text{K}} = 20.9 + 5.9z + 0.5z^2
\end{equation}
We rewrite this slightly with a free parameter describing the polynomial scaling of the kinetic temperature with redshift:
\begin{equation}
    T_k(z) = 3.5\text{\,K} + T_{k,2}(5.9 + z)^2,
\end{equation}
with $T_{k,2}=0.5$\,K being the fiducial value.

The molecular cloud count per halo in the original Model B was adjusted to yield a reasonable if optimistic cosmic molecular gas mass density at $z\sim7$, so extrapolating to lower redshift will require accounting for the exhaustion of molecular gas in recent cosmic epochs. We thus introduce a power law factor into the functional form of~\cite{COMAPTDX}, adding two free parameters to our model:
\begin{equation}N_\text{MC}(M_h) = \frac{1.9\times10^{7}}{1+(10^{12.3}\,M_\odot/M_h)^{3/2}}\times\delta_\text{MC}(1+z)^{\gamma_\text{MC}}.\end{equation}
The fiducial values are $\delta_\text{MC}=2.2\times10^{-3}$ and $\gamma_\text{MC}=3$; this leads to the power law factor being $\approx1$ at $z\sim7$, thus recovering the fiducial Model B cloud counts at late reionisation.

Finally, we must define one more input for Radex, the CO column density given halo mass and redshift:
\begin{equation}N_\text{CO} = 5.2\times10^{16}\times n_\text{H2}^{2/3}Z'(M_h)/Z_\odot.\end{equation}
Here we assume the gas-phase metallicity $Z'(M_h)$ follows a double power law of fixed form, varying only the overall normalisation:
\begin{equation}
    \frac{Z'(M_h)}{Z_\odot} = \frac{Z'_\text{break}}{1+(10^{12.3}\,M_\odot/M_h)^{1/3}},\end{equation}
with a fiducial value of $Z'_\text{break}=2$ such that $Z'(M_h=10^{12.3}\,M_\odot)=1\,Z_\odot$, with relatively weak evolution against halo mass.

Multiplying the CO emitting area per cloud $\pi r_\text{MC}^2$ by the velocity-integrated flux $F_{J,\text{MC}}$ obtained from Radex (against a background temperature of $(1+z)T_\text{CMB}=2.7255(1+z)$\,K), we obtain the CO($J\to J-1$) luminosity per cloud and scale this by $N_\text{MC}(M_h)$ to obtain the total luminosity per halo:
\begin{equation}
    L'_J(M_h)=\pi r_\text{MC}^2F_{J,\text{MC}}N_\text{MC}(M_h).
\end{equation}
Based on the assumed halo mass function $dn/dM$, the integrated global CO line intensity is given by
\begin{equation}
    \avg{I}_J=\frac{2k_B\nu_J^2}{c^2H(z)}\int dM_h\,\frac{dn}{dM_h}\,L'_J(M_h),
\end{equation}
where $\nu_J\approx115.27\times J$\,GHz is the rest-frame emission frequency of the corresponding CO line.
\begin{figure}
    \centering
    \includegraphics[width=0.96\linewidth]{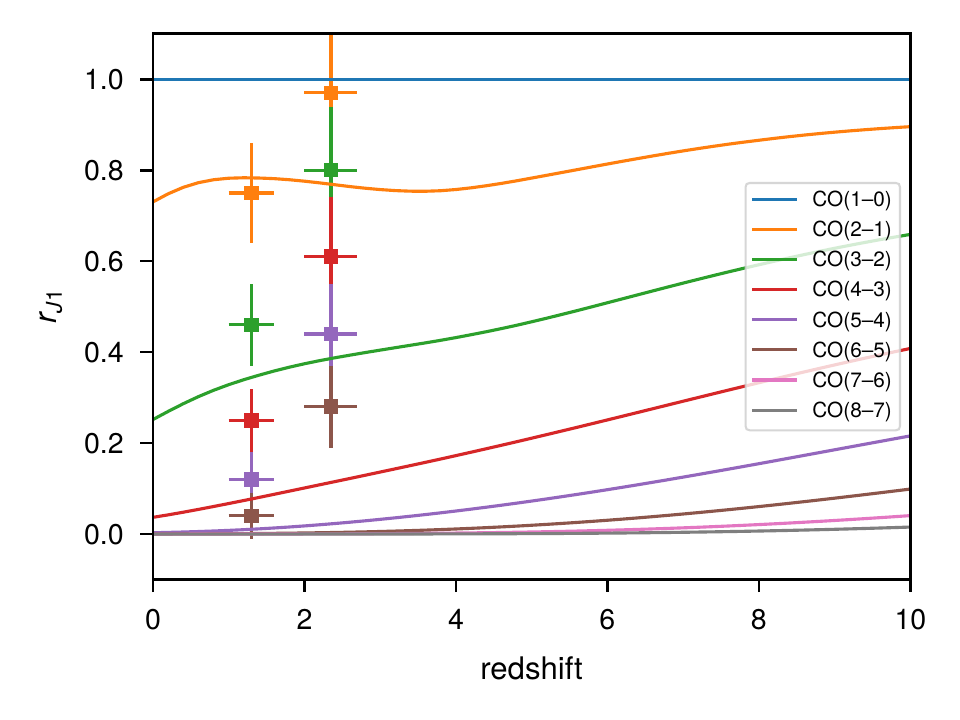}
    \includegraphics[width=0.98\linewidth]{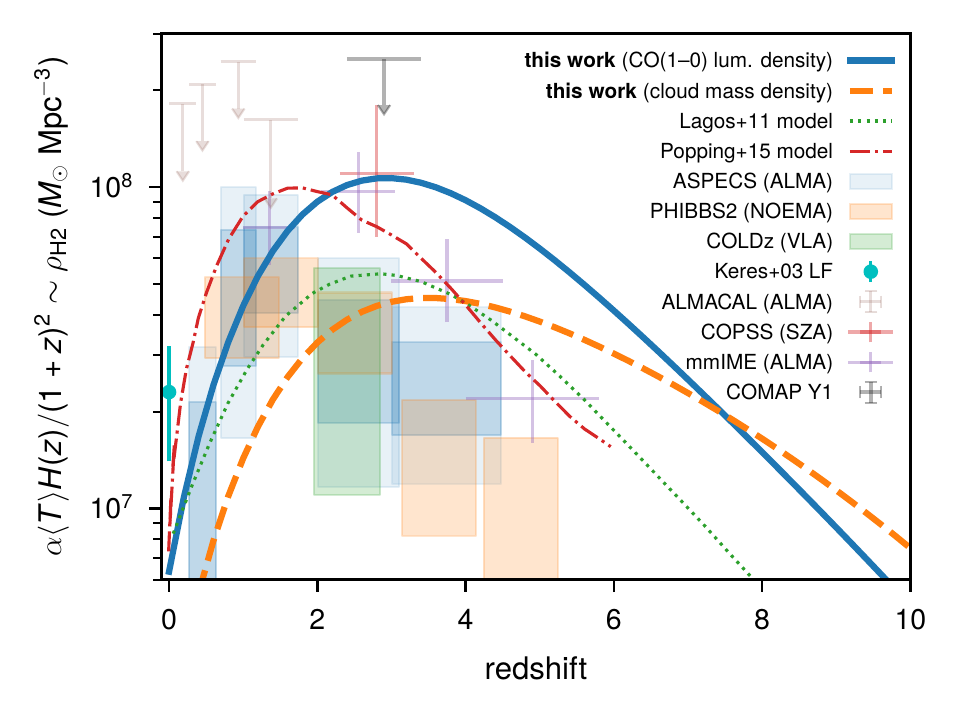}
    \caption{Redshift evolution of CO emission across the multiple rotational transitions modelled in this work. \emph{Upper panel:} Global line temperature ratios $r_{J1}=\avg{I}_J/(J^2\avg{I}_1)$ across redshift according to our model. For comparison, we show CO line ratios modelled at $z\sim1$--3 based on ASPECS LP data~\citep{Boogaard20}. \emph{Lower panel:} Molecular gas density, either theoretically modelled~\citep{Lagos11,Popping15} or inferred from CO abundances measured by local volume observations~\citep{Keres03}, high-redshift interferometric line emitter surveys~\citep{Decarli20,PHIBBS2Lenkic,COLDzLF}, high-redshift absorption surveys~\citep{Klitsch19}, and CO LIM surveys~\citep{Keating16,mmIME-ACA,COMAPESV}. For the model primarily considered in our work, we show both the theoretical molecular gas density and the density that would be recovered from the cosmic average CO(1--0) line intensity.}
    \label{fig:COcheck}
\end{figure}

We show the global line temperature ratios $r_{J1}=\avg{I}_J/(J^2\avg{I}_1)$ in the upper panel of~\cref{fig:COcheck}. For comparison, we show line ratios modelled by~\cite{Boogaard20}, which are mostly similar at $z<2$ for lower energy transitions but are in tension at $z>2$. The \cite{Boogaard20} analysis however is based on modelling the excitation of individual objects selected based on CO line luminosities (with $J>1$), and so may be more skewed towards high-excitation environments.

The lower panel of~\cref{fig:COcheck} plots $\avg{I}_1$ cast as a molecular gas mass density, assuming a CO--H$_2$ conversion of $\alpha_\text{CO}=3.6\,M_\odot\,($K\,km\,s$^{-1}$\,pc$^{-2})^{-1}$, compared to other CO observational constraints using the same or similar conversions. In particular, we show constraints from the ASPECS LP~\citep{Decarli20}, COLDz~\citep{COLDz,COLDzLF}, and PHIBBS2~\citep{PHIBBS2,PHIBBS2Lenkic} untargeted CO emission surveys, the ALMACAL CO absorption survey~\citep{Klitsch19}, the COPSS~\citep{Keating16} and mmIME~\citep{mmIME-ACA} interferometric LIM surveys, and the Early Science result from the CO Mapping Array Project (COMAP) Pathfinder single-dish LIM survey~\citep{COMAPESV}, as well as a constraint at $z\sim0$ from~\cite{Keres03} obtained by integrating a local volume luminosity function. Our model shows good agreement with CO observations at $z<2$ and preference for the LIM-based constraints at $z\sim3$. Predictions at $z>3$ skew optimistic but are by no means strongly excluded, given the range spanned by observational constraints at higher redshifts.

For reference, we also show the actual underlying molecular gas mass density, based on integrating the total mass of molecular clouds sourcing the CO emission in the model. Comparing the two curves from our model shows that while $\alpha_\text{CO}=3.6\,M_\odot\,($K\,km\,s$^{-1}$\,pc$^{-2})^{-1}$ is conventional, lower values may be more appropriate at high redshift \citep[cf.][]{Breysse22mmIME}. We also show a few simulation-based models from the literature~\citep{Lagos11,Popping15} for comparison, showing that our model is within a factor of order unity of these predictions for the most part.

\subsection{Ionized carbon line emission model}
\label{sec:CIImodel}
We extrapolate the \cii{} model of \cite{CII_MHIZ}, a model of \cii{} as a tracer of H\,\textsc{i} mass in galaxies (as used by, e.g.,~\citep{Heintz_2021,Heintz22}), calibrated against expectations based on post-processing of the Feedback in Realistic Environments (FIRE) simulations by~\cite{liang2023}. Broadly speaking, we expect the majority of \cii{} emission to originate from the interface between neutral and ionized gas, modulated by the interstellar radiation field (ISRF) and the gas-phase metallicity. As a result, we expect the \cii{} luminosity in a given halo to be proportional to the H\,\textsc{i} mass $M_\text{HI}$ and the gas-phase metallicity $Z'(M_h)$. Depending on the strength and morphology of the ISRF relative to shielding mechanisms, the fraction of gas emitting in \cii{} may not be 100\%. In~\cite{CII_MHIZ} we implicitly assumed a value of $f_\text{\cii{}}=1$ for this fraction, which is appropriate for the low metallicity, high redshift environments that were the focus of that work. In this work, however, we will assume some evolution of $f_\text{\cii{}}$ with both halo mass and redshift, so that
\begin{equation}
    L_\text{\cii{}}(M_h,z) = \alpha_\text{\cii{}}f_\text{\cii{}}(M_h,z)M_\text{HI}(M_h,z)\frac{Z'(M_h)}{Z_\odot}.
\end{equation}
We fix $\alpha_\text{\cii{}}$ at its fiducial value of $0.024\,L_\odot\,M_\odot^{-1}$ used in~\cite{CII_MHIZ}. To calculate the \textsc{H\,i} gas mass per halo, we use the functional form of~\cite{paco_2018}, which follows a power law with an exponential cutoff at low halo mass:
\begin{align}
    \frac{M_\text{HI}(M_h,z)}{M_\odot}&= M_0 \left(\frac{M_h/M_\odot}{M_{\rm{min}}}\right)^{\alpha_\text{HI}}\nonumber\\&\hspace{1.86cm}\times\exp{\left[-\left(\frac{M_{\rm{min}}}{M_h/M_\odot}\right)^{0.35}\right]}.
    \label{paco_eq}
\end{align}
Given the relative lack of evolution of the parameters $\alpha_\text{HI}$, $M_0$, and $M_\text{min}$ at high redshift, we simply used the values given by~\cite{paco_2018} at $z=5$ scaled down by 70\%. Here, we use smooth approximations to the parameter values across all redshifts:
\begin{align}
    \alpha_\text{HI}(z) &= 0.79-\frac{0.55}{1+z}+0.13\exp{\left(-\frac{(z-3.6)^2}{2\times0.82^2}\right)};\\
    M_0(z) &= 1.5\times10^9+4.2\times10^{10}\times e^{-z/0.68}\nonumber\\&\hspace{1.86cm}+10^{10}\exp\left(-\frac{(z-1.8)^2}{2\times0.54^2}\right);\\
    M_\text{min}(z) &= 1.4\times10^{10}+2\times10^{12}\times e^{-z/0.73}\nonumber\\&\hspace{1.86cm}+2.3\times10^{11}\exp\left(-\frac{(z-1.8)^2}{2\times0.54^2}\right).
\end{align}
The Gaussian additive terms reflect the bumpy evolution of cosmic \textsc{H\,i} abundance seen in both observations and simulations.

We assume that the \cii{}-emitting gas fraction relates to both metallicity and the molecular cloud population. The first is clear since the absolute amount of \cii{}-emitting gas begins to saturate in high metallicity systems~\citep{liang2023}. The connection to the molecular cloud population is less direct, but the late-epoch decline in molecular gas abundance should relate to a decline in the strength of the ISRF and thus the amount of gas available for excitation at the \textsc{H\,i}--\textsc{H\,ii} interface. Thus, we consider this entirely empirical form for the \cii{}-emitting gas fraction:
\begin{align}
    f_\text{\cii{}}(M_h,z) &= \left[1+0.1Z'(M_h)^2\right]^{-1}\nonumber\\&\hspace{1.86cm}\times\left[1+5(1+z)^{-\gamma_\text{MC}}\right]^{-1}.
\end{align}
This form asymptotes to 1 as metallicity decreases or (for $\gamma_\text{MC}>0$) redshift increases.

\begin{figure}
    \centering
    \includegraphics[width=0.96\linewidth]{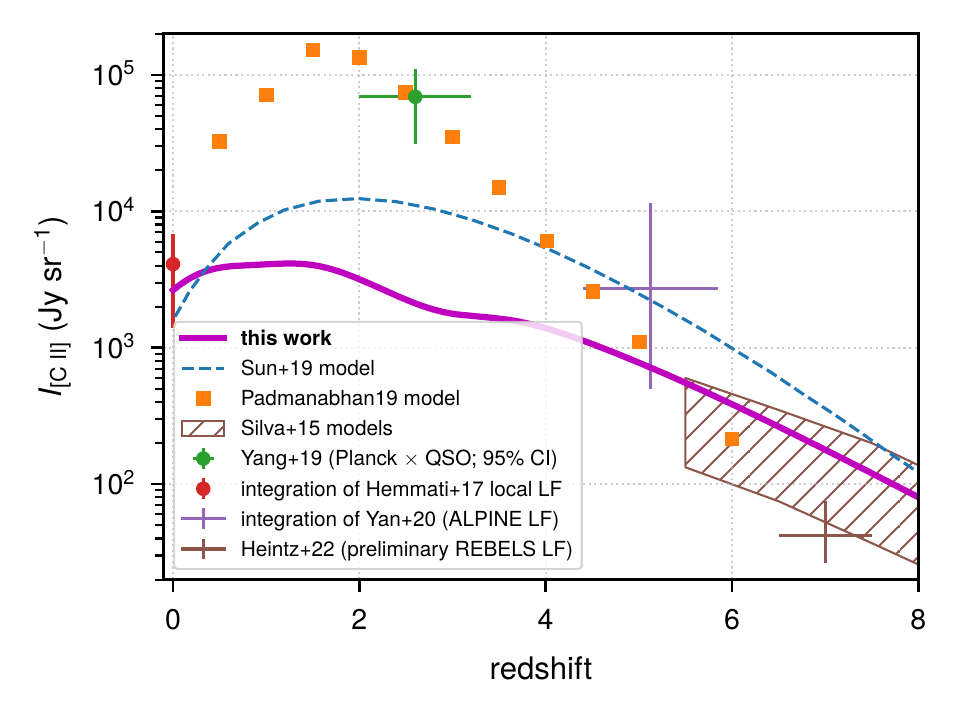}
    \caption{Redshift evolution of the cosmic average \cii{} line intensity, considered by models~\citep{Silva15,Padmanabhan19,Sun19} and constrained by observations~\citep{Hemmati17,ALPINELF,Heintz22}, as indicated in the legend.}
    \label{fig:CIIcheck}
\end{figure}

Note that in defining this model, we are using the same $Z'(M_h)$ function and the same $\gamma_\text{MC}$ parameter as used in the CO model, for physical consistency. The redshift evolution in particular allows us to come into reasonable agreement with local constraints from~\cite{Hemmati17}, represented as an integrated cosmic average \cii{} intensity in~\cref{fig:CIIcheck} as calculated by~\cite{Anderson22}.

Considering other models and constraints, the current model is in reasonable agreement with reality and other high-redshift models. Our model does not predict intensities at $z\sim2$ as high as some other models in the literature, including that of~\cite{Padmanabhan19}, which anchors itself to the anomalous excess detected by~\cite{Yang19} in cross-correlating Planck intensity maps with quasar catalogues (\replaced{c.f.}{cf.}, also,~\citep{Pullen18,Switzer19}). Our model may still be justified, however, as alternative explanations for this excess do exist and have yet to be excluded. Our model also falls in the middle of the pack relative to models like that of~\cite{Silva15} or~\cite{Sun19} at $z\gtrsim6$, as may be expected for a model originally tuned to forecast late-reionisation \cii{} LIM signals. Current observational constraints at these redshifts derive from targeted surveys like ALPINE~\citep{ALPINELF} and REBELS~\citep{Heintz22}. Determination of the luminosity functions are fairly loose at time of writing, and thus our model is not in ruinous tension with the resulting constraints.\footnote{Of these constraints, we calculate the constraint implied by ALPINE ourselves using their Schechter fitting parameters and errors and integrating down to the lowest edge of the lowest luminosity bin considered, $L_\text{\cii{}}=10^{7.5}\,M_\odot$.}

\begin{figure}
    \centering
    \includegraphics[width=0.986\linewidth]{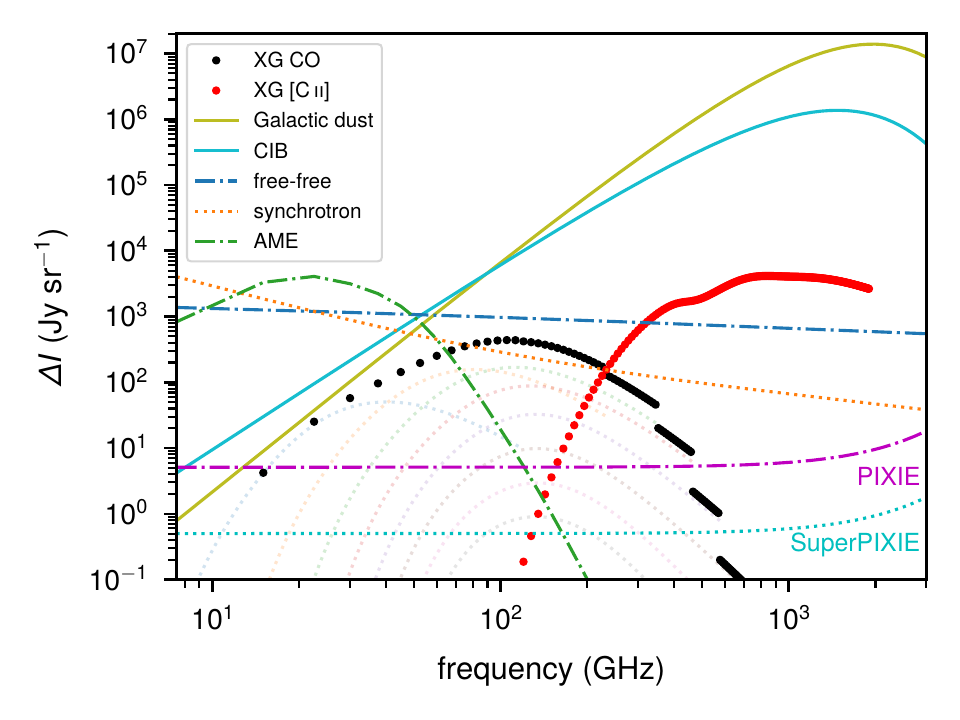}
    
    \includegraphics[width=0.986\linewidth]{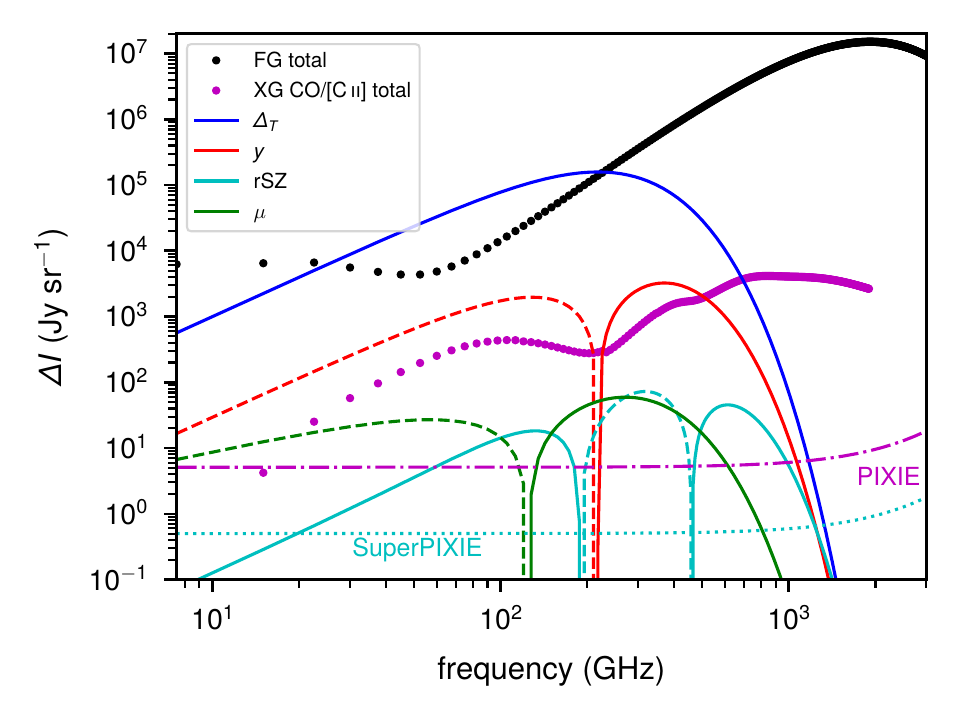}
    \caption{Components of the sky-averaged intensity spectrum at frequencies from 7.5 GHz to 3 THz. \emph{Upper panel:} Foreground components of the spectrum, including the extragalactic (`XG') CO and \cii{} components modelled anew in this work as well as the other components considered in the main text. We also show the contribution of individual CO transitions to the total extragalactic CO emission (as very faint dotted lines). \emph{Lower panel:} Cosmological spectral distortions of interest to a \PIXIE-like survey as outlined in the main text, compared to the total foreground spectrum (black dots) and the extragalactic carbonic contribution (magenta dots). Both panels also show the expected sky-averaged sensitivity of the \PIXIE and \SPIXIE scenarios outlined in the main text.}
    \label{fig:breakdown}
\end{figure}

Before moving to discussing forecasting constraints on the CO and \cii{} global signals, we summarise the end result of our models graphically in~\cref{fig:breakdown}, including models described in~\cite{Abitbol17} beyond the scope of original modelling in this work. The shape of the final CO global signal is closer to the result of Serra, Dor\'{e}, and Lagache~\cite{Serra16} than to that of Mashian, Loeb, and Sternberg~\cite{Mashian16}, with distinct steps up in intensity as one scans from high to low frequencies and sees additional CO transitions enter starting at $z=0$. Given that our assumptions around CO excitation match the $z\sim1$ results of~\cite{Boogaard20} fairly well (particularly for the lowest energy transitions), the lack of such a `staircase' or `sawtooth' feature in the prediction of~\cite{Mashian16} suggests under-prediction of CO excitation at low redshift. A potential source of \changeJfirstround{uncertainty} in predicting CO excitation is the choice of~\cite{Mashian15,Mashian16} to model the molecular gas content of each halo as a single cloud, rather than a collection of many smaller clouds as in our model and in the original Model B of~\cite{COMAPTDX}.

Despite this, our CO modelling approach otherwise ties closer to that of~\cite{Mashian16}, applying radiative transfer modelling on top of scaling relations around molecular cloud regions, than that of~\cite{Serra16}, an empirical modelling approach that connects all CO lines in uncorrelated fashion to the IR luminosities of galaxies. In this context there is a strong argument for preferring the former, with its intrinsically physically consistent modelling of the CO spectral line energy distribution. \changeJfirstround{This also directly highlights the exciting opportunities of future CMB spectrometers to shed light on the underlying physical modeling of the average cosmic line signals and their fluctuations at the largest scales.}

\subsection{Fisher forecasting}
\label{sec:fishermethods}
The primary purpose of this work is not to optimise analyses of spectral distortion data for expected sensitivities of future experiments, but rather to illustrate the relative penalty to constraining power on $y$- and $\mu$-type distortions incurred by CO/\cii{} foreground spectra. Given this scope, a Fisher forecast is sufficient as opposed to a more detailed end-to-end mock analysis.

In general, the Fisher matrix for an observable vector ${O}_\alpha$ with covariance matrix $\mathbf{C}_{\alpha\beta}$ is given by
\begin{equation}
    \mathbf{F}_{rs} = \sum_{\alpha,\beta}\frac{\partial{{O}}_{\alpha}}{\partial\lambda_r}\mathbf{C}^{-1}_{\alpha\beta}\frac{\partial{{O}}_{\beta}}{\partial\lambda_s},\label{eq:fisher}
\end{equation}
The inverse of this yields the parameter covariance matrix, the diagonal of which gives the parameter variances $\{\sigma_\lambda^2\}$.

Here the observable vector is simply the observed deviation $\Delta I$ of the spectrum from the expected CMB monopole, in channels of 15 GHz bandwidth ranging from 15 GHz to 3 THz in central frequency. We approximate the derivatives with central difference quotients obtained using parameter values 5\% above and below the fiducial value for each parameter.

The non-CO/\cii{} parameters\replaced{allowed to vary (and their fiducial values)}{, most of which we allow to vary at some point,} are as follows\added{ (with fiducial values also disclosed)}:
\begin{itemize}
    \item $\Delta_T=1.2\times10^{-4}$, \changeJfirstround{an assumed} relative deviation away from the assumed $T_\text{CMB}=2.7255$\,K;
    \item $y=1.77\times10^{-6}$, the integrated parameter describing the cumulative non-relativistic Compton-$y$ distortion;
    \item $kT_\text{eSZ}=1.282$\,keV and $\omega_2^\text{eSZ}=1.152$, describing the first and second moments of the $y$-weighted electron temperature distribution, describing the relativistic Compton-$y$ distortion; 
    \item $\mu=2\times10^{-7}$, the integrated parameter describing the primordial $\mu$-type distortion;
    \item $A_D=1.36$ MJy sr$^{-1}$, $\beta_D=1.53$, and $T_D=21$\,K, the parameters describing the modified blackbody spectrum of Galactic dust;
    \item $A_\text{CIB}=0.346$ MJy sr$^{-1}$, $\beta_\text{CIB}=0.86$, and $T_\text{CIB}=18.8$\,K, the parameters describing the modified blackbody spectrum integrated across the cosmic infrared background;
    \item $A_S=288$ Jy sr$^{-1}$, $\alpha_S=-0.82$, and $\omega_S=0.2$, parameters describing the amplitude and shape of the Galactic synchrotron emission spectrum;
    \item $A_{FF}=300$ Jy sr$^{-1}$, describing the amplitude of the Galactic free-free emission spectrum;
    \item and $A_\text{AME}=1$, describing the relative amplitude of the AME (spinning dust) spectrum.
\end{itemize}
\added{We refer the reader to the earlier study of~\cite{Abitbol17} for the parameterisations of the modified blackbody, synchrotron, and free-free spectra. In particular, for the synchrotron and free-free components, we fix the characteristic frequencies and temperatures at the same values as~\cite{Abitbol17} did, namely a knee frequency of $\nu_0=100$\,GHz for synchrotron emission, and an electron temperature and knee frequency of $\{T_e,\nu_\text{FF}\}=\{7000$\,K$,255.33$\,GHz$\}$ for free-free emission.}

\changeJfirstround{Following earlier studies \citep{Abitbol17, Rotti2021, Chluba2021Voyage}, we impose conservative Gaussian priors on $A_S$ and $\alpha_S$ with $\sigma$ set to 10\% of the fiducial values. \changeJ{Anticipating major advancements in ground-based low-frequency surveys \citep[e.g.,][]{QUIJOTE2023I, ELFS2023}, one can expect this to be significantly improved}}.

Note that the fiducial value of $\mu$ we assume is ten times higher than the fiducial value assumed for the work of~\cite{Abitbol17}, which based its value of $\mu=2\times10^{-8}$ on predictions from standard \changeJ{cosmology~\citep[e.g.,][]{Chluba16}. In a Fisher treatment, the precise fiducial value does not affect the forecast significantly. In addition, there are a wide range of viable models \citep[e.g.,][]{Chluba2013fore} that can produce a value of $\mu$ even up to the {\it COBE/FIRAS} limit. In reality, recent discussions of primordial black holes and the detection of a stochastic gravitational wave background by NANOGrav suggest new physics possibly linking to enhanced small-scale power \citep{NANOGravDetection2023, NANOGravSMBHB2023}. Related scenarios could produce a $\mu$ distortion well in excess of the standard $\Lambda$CDM expectation \citep{Chluba2012inflaton, Cyr2024power}, which itself relies on an extrapolation over a vast range of scales, justifying our choice of reference. We also stress that, after all,} 
the focus of this work is not a robust prediction of primordial spectral distortions, but rather an examination of CO and \cii{} emission in spectral distortion experiments. As the $\mu$-type distortion in our calculation is given by an overall scaling of a fixed spectral template by the parameter $\mu$, the fiducial value assumed for $\mu$ has no effect on a Fisher analysis of the kind carried out in this work, dependent solely on derivatives of the observable vector.

We may add consideration of CO and \cii{} parameters in one of two different ways.
\begin{itemize}
    \item We assume fixed templates for the shape of the CO and \cii{} global signals, and allow each of the components (total CO or \cii{}) to vary up or down with dimensionless scaling factors $A_\text{CO}$ and $A_\text{\cii{}}$, both with fiducial values of 1. This is the approach taken by~\cite{Abitbol17} in modelling the extragalactic CO foreground.
    \item We take four of the parameters introduced in~\autoref{sec:COmodel} and~\autoref{sec:CIImodel}, allowing the CO and \cii{} contributions to vary more freely in both amplitude and shape. The four parameters in this context and their fiducial values are $Z'_\text{break}=2$ (controlling the metallicity), $T_{k,2}=0.5$\,K (controlling the redshift evolution of kinetic temperature), $\gamma_\text{MC}=3$ (controlling the redshift evolution of cloud count per halo), and $\delta_\text{MC}=2.2\times10^{-3}$ (controlling the overall normalisation of cloud count per halo).
\end{itemize}

\begin{figure}
    \centering
    \includegraphics[width=0.96\linewidth]{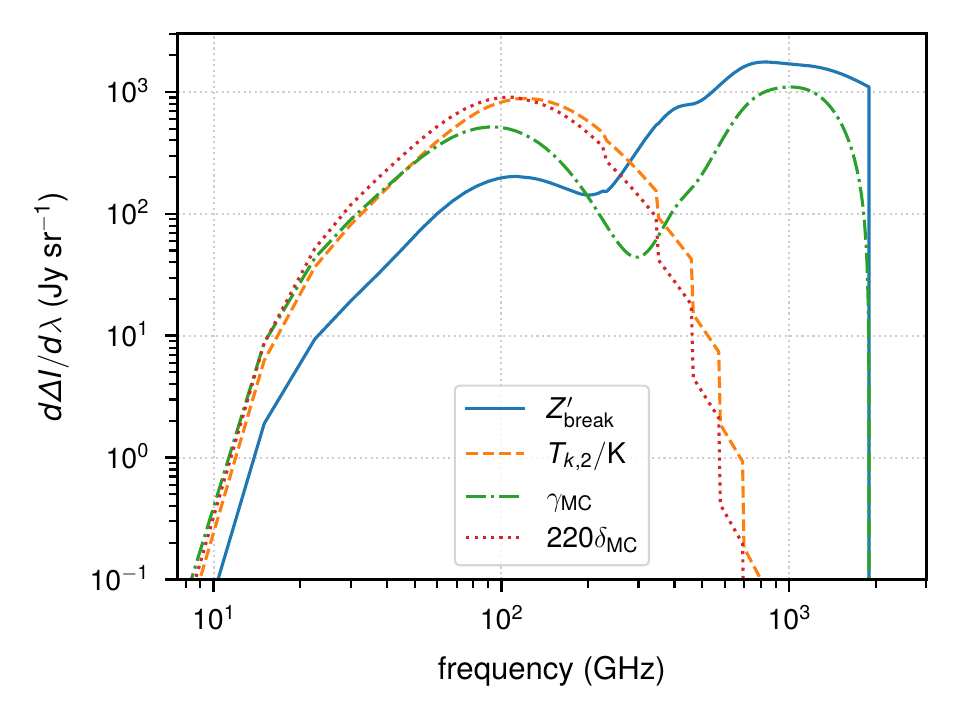}
    \caption{Derivatives of the sky-averaged intensity spectrum against the parameters governing the CO and \cii{} spectra, as used for Fisher forecasting in this work. We scale the derivative against $\delta_\text{MC}$ by a factor of 220 to show it at roughly the same scale as the other derivatives.}
    \label{fig:derivatives}
\end{figure}

\begin{figure}
    \centering
    \includegraphics[width=0.96\linewidth]{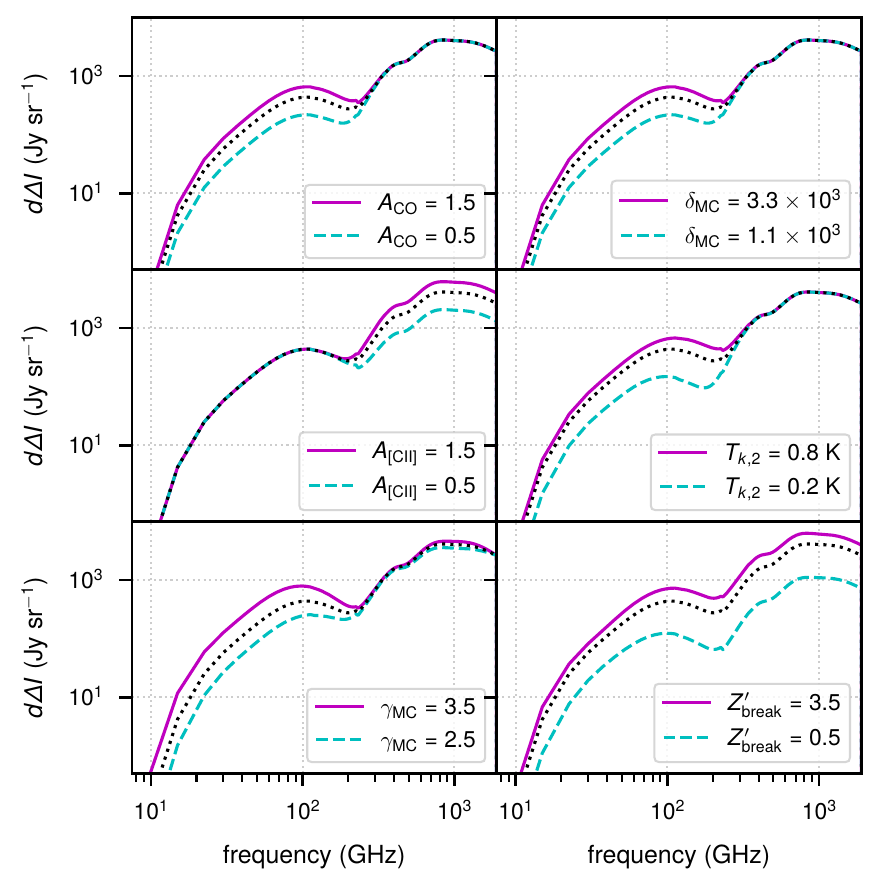}
    \caption{Illustration of variations in the CO/\cii{} global signal due to changes in either an overall scaling of the CO and \cii{} templates ($A_\text{CO}$ and $A_\text{\cii{}}$) or one of the parameters in our four-parameter model for the CO/\cii{} global signal. The fiducial prediction (dotted) is shown in each panel, with the fiducial value for each parameter lying at the linear midpoint of the range of variation in each case.}
    \label{fig:d_DeltaI}
\end{figure}

\autoref{fig:derivatives} shows derivatives of the observable vector with respect to the four parameters governing the CO and \cii{} components in the second approach. We also show the total CO/\cii{} global signal under these variations in~\cref{fig:d_DeltaI}, along with simple overall scalings of either one of the CO and \cii{} components. Whatever the presentation, it is evident that $Z'_\text{break}$ controls the overall normalisation of the combined CO-plus-\cii{} signal, as both components scale roughly linearly with metallicity. It is also evident that $\delta_\text{MC}$ controls the overall normalisation of the total CO component alone, as expected given that it uniformly scales the cloud count per halo.

\addedtwo{\begin{figure}
    \centering
    \includegraphics[width=0.986\linewidth]{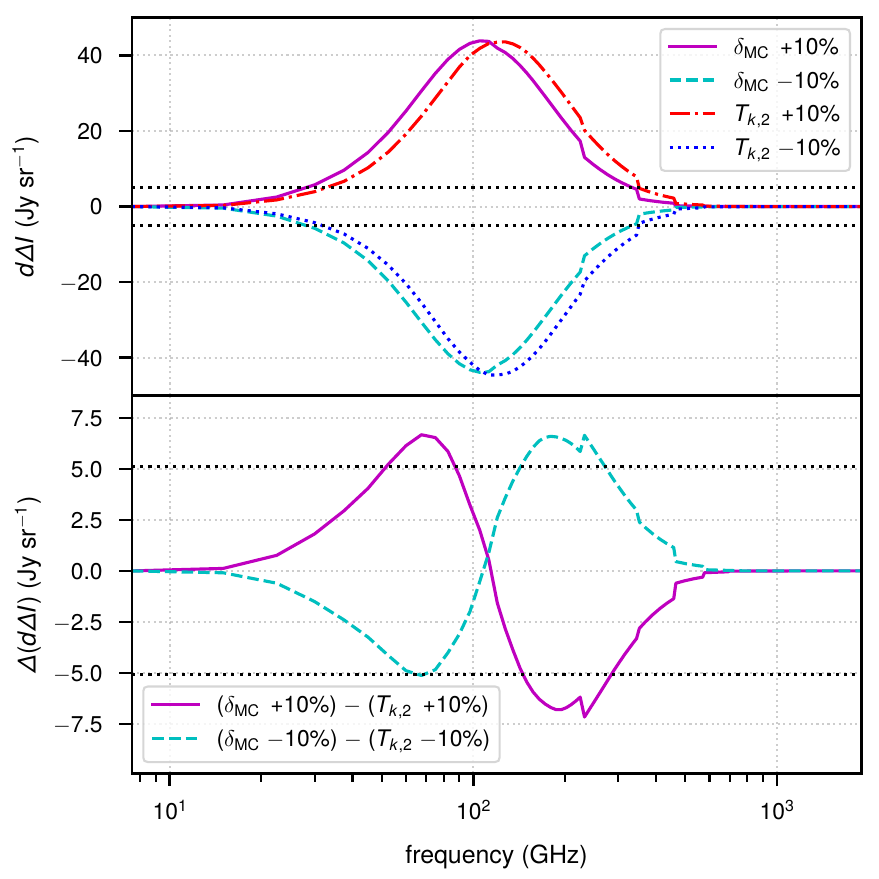}
    \caption{Illustration of the distinct effects of $\delta_\text{MC}$ and $T_{k,2}$ on the CO global signal. \emph{Upper panel:} Changes from the fiducial prediction introduced by changing either parameter by $\pm10\%$ as indicated in the legend. \emph{Lower panel:} Differences between the changes calculated in the upper panel, showing how a 10\% change in one parameter differs from a 10\% change in the other. In both panels, we represent the noise level of \PIXIE as black dotted horizontal lines at $\pm5.1$\,Jy\,sr$^{-1}$; note that the \SPIXIE concept would achieve a mission sensitivity improved by a factor of ten.}
    \label{fig:Delta_d_DeltaI}
\end{figure}}

It is worth remarking on the effect of the remaining two parameters. The effect of increasing $\gamma_\text{MC}$ is overall to increase the CO and \cii{} global signals, but especially the $z\gtrsim1$ segment. This is easy to see based on how in~\cref{fig:derivatives}, changing $\gamma_\text{MC}$ does not introduce the `staircase' steps in intensity associated with the $z=0$ intensity of individual lines, which changing $\delta_\text{MC}$ does for example. Given our parameterisation, this behaviour makes sense both for CO (where the cloud count will grow with a higher power-law index) and for \cii{} (where the redshift-dependent factor in $f_\text{\cii{}}(M_h,z)$ will approach unity more quickly). Meanwhile, the effect of increasing $T_{k,2}$ is to increase excitation of higher-$J$ CO lines, which naturally results in a skewed increase in the CO signal favouring the higher-frequency part of the spectrum. The effect is somewhat degenerate with the effect of $\delta_\text{MC}$ (i.e., an overall scaling of the CO signal), but still distinguishable in principle \addedtwo{as we illustrate in~\autoref{fig:Delta_d_DeltaI} and }as the Fisher forecast will show.

We are still missing one element of the Fisher forecast, however, which is the covariance matrix for our observable. We consider two benchmark concepts for a microwave spectrometer.
\begin{itemize}
    \item The Primordial Inflation Explorer (\PIXIE;~\citep{Kogut11, Kogut2016SPIE}) is a proposed instrument that includes absolute intensity spectrometry of the microwave sky. Based on the noise level of $2.4\times10^{-22}$ W m$^{-2}$ sr$^{-1}$ Hz$^{-1}$ quoted for a one-second integration, a survey period of one year (365.25 days), and 70\% sky coverage, we obtain an expected noise level of $\sigma_I=5.1$ Jy sr$^{-1}$.

    \item The \changeJfirstround{works of~\cite{Chluba2021Voyage, V2050}}, written for consideration in the ESA `Voyage 2050' long term planning study, considers the possibility of a flagship-class mission with duplicated spectrometer modules collecting useful data for 70\% of the sky across six years. While the fiducial concept would make use of a cold Fourier-transform spectrometer (FTS), future advances in on-chip spectrometers may reduce mission complexity.

    In this spirit of this `Voyage 2050' roadmap, we consider an \changeJfirstround{enhanced} \SPIXIE concept where the \PIXIE instrument is effectively duplicated and operated sufficiently to achieve a mission sensitivity of $\sigma_I=0.5$ Jy sr$^{-1}$, a figure on the order of mission sensitivities quoted in~\cite{V2050}.
\end{itemize}
In both cases, we assume a low-pass filter profile based loosely on the description of~\cite{Kogut11}, such that the actual covariance matrix is a diagonal matrix proportional to the Kronecker delta $\delta_{\alpha\beta}$, and given by
\begin{equation}
    \mathbf{C}_{\alpha\beta}=\delta_{\alpha\beta}\sigma_I\left(1+\frac{\nu}{5\,\text{THz}}\right)^3\left(1+\frac{\nu}{3\,\text{THz}}\right)^{1/2},
\end{equation}
effectively supposing the presence of six identical single-pole low-pass filters with a knee wavelength of 60 microns and one separate low-pass filter with a knee wavelength of 100 microns.\footnote{This supposition differs from the outline of~\cite{Kogut11}, which includes five filters with a knee wavelength of 60 microns. However, one of the 60 micron `filters' (actually the absorption profile of the bolometers) could be distinct from the other four (actually the wire grids of the optical polarisers), and effectively similar to two cascaded single-pole filters.} The \PIXIE sensitivity curve shown in~\cref{fig:breakdown} thus obtained is a reasonable match to the \PIXIE sensitivity curve given by~\cite{Abitbol17}.\addedtwo{ We do not forecast for the potential use of frequency channels at $\nu>3$ THz, given that this would require further refinement in models of both the instrument and Galactic or extragalactic dust at these super-terahertz frequencies, which is well beyond the scope of the present exploratory work.}

\begin{figure}
    \centering
    \includegraphics[width=0.96\linewidth]{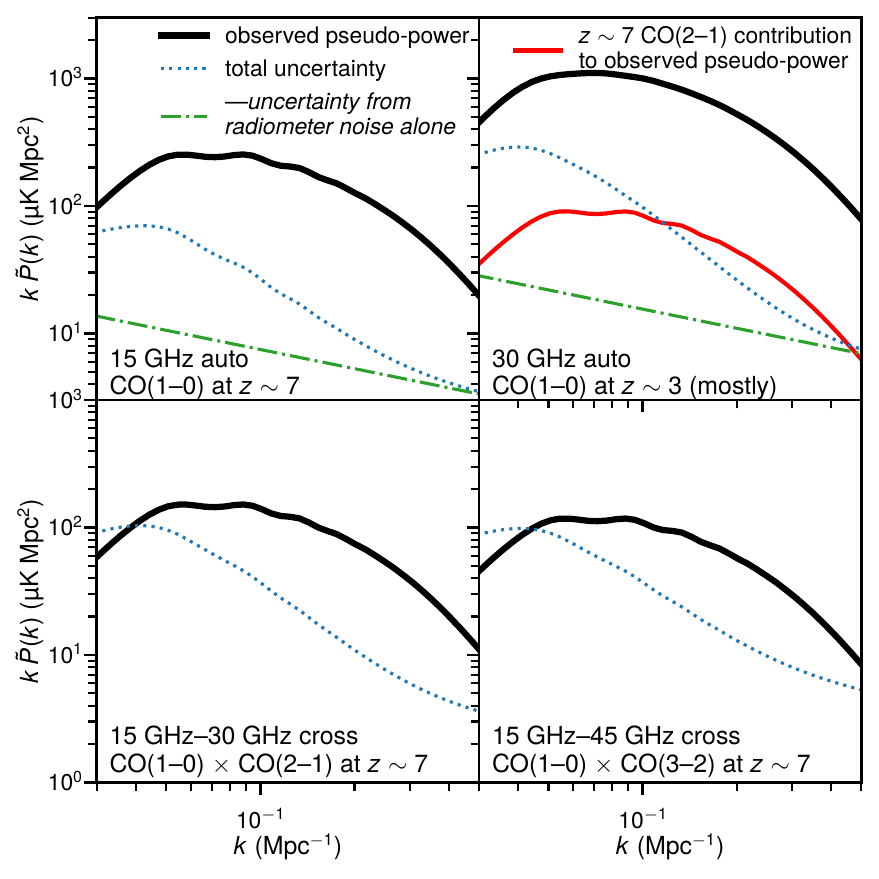}
    \caption{Predictions for CO auto and cross pseudo-power spectra (i.e., power spectra including observational distortions) and sensitivities, given the COMAP `Triple Deluxe' concept and our fiducial extragalactic CO line emission model. Total uncertainties take variance from interloper emission into account, and are evaluated for $k$-bins of size $\Delta[\log{(k\,\text{Mpc})}]=0.03$.}
    \label{fig:COLIM_Pk}
\end{figure}

In some cases we also consider a CO LIM experiment, specifically the COMAP `Triple Deluxe' concept considered by~\cite{COMAPTDX}. We assume the experiment is capable of measuring CO emission around observer-frame frequencies of 15 GHz, 30 GHz, and 45 GHz, and that reliable measurements are possible of 15 GHz and 30 GHz auto power spectra due to extragalactic CO, along with 15 GHz--30 GHz and 15 GHz--45 GHz cross power spectra. In particular, these observables shown in~\cref{fig:COLIM_Pk} will isolate CO(1--0) emission from $z\sim3$ as well as CO(1--0) auto, CO(1--0)--CO(2--1) cross, and CO(1--0)--CO(3--2) cross power at $z\sim7$. We derive power spectrum values and covariances following the formalism of~\cite{COMAPTDX}, resulting in a total detection significance of 106 across all observables considered. We assume no covariance between the CO LIM and spectral distortion measurements in this work, and furthermore we do not consider the possibility of cross-correlating the CO LIM and \PIXIE (or \SPIXIE) data sets in any way. The forecasts in this work thus do not represent the limit of what can be achieved with the combination of the two experiments, let alone LIM and absolute spectrometer experiments in general.

\section{Results and discussion}
\label{sec:results}
With our models and Fisher forecasting formalism, we may consider the CO and \cii{} contributions to the sky-averaged microwave spectrum from two viewpoints. First, we consider in~\autoref{sec:res_sg} how we can use the CO and \cii{} signals to learn about conditions in the interstellar medium across a wide range of redshifts. We then consider in~\autoref{sec:res_fg} the necessity of accounting for the CO and \cii{} line emission through nuisance foreground parameters in claiming detection of other cosmological signals.
\subsection{Carbonic lines as signals: constraining ISM conditions through cosmic time}
\label{sec:res_sg}
\begin{figure}
    \centering
    \includegraphics[width=0.96\linewidth]{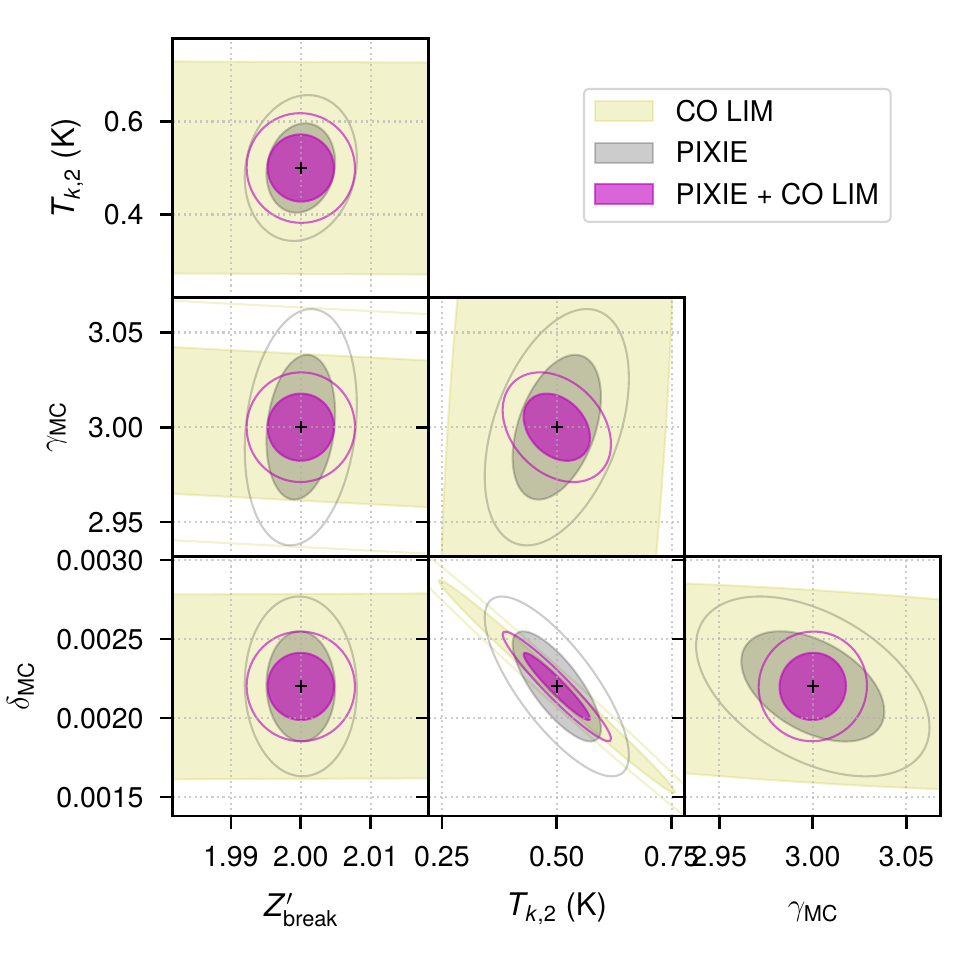}
    \caption{68\% (filled) and 95\% (unfilled) confidence ellipses for the parameters of the CO/\cii{} model derived from Fisher forecasts using CO LIM alone (yellow), \PIXIE alone (grey), or both data sets (magenta). Note that the joint \PIXIE-plus-CO LIM forecast does not include any potential direct cross-correlation between CO LIM data and \PIXIE data.}
    \label{fig:ellipses}
\end{figure}
In considering the constraints on the CO and \cii{} signals provided by absolute microwave spectrometry, and what the constraints translate to in scientific terms, it is best to consider the full four-parameter model. However, given the high uncertainty in the brightness of the line emission components, even an overall detection of their amplitudes may be interesting. To this end, we first present the simplest case where we scale fixed templates for CO and \cii{} by dimensionless factors $A_\text{CO}$ and $A_\text{\cii{}}$. In this initial Fisher forecast, we fix the parameters governing the relativistic corrections to the $y$-type distortion ($kT_{\rm eSZ}$ and $\omega_2^{\rm eSZ}$), and assume \PIXIE-like sensitivities.

\begin{table}
    \centering
    \begin{ruledtabular}\begin{tabular}{rcccc}
        & \multicolumn{4}{c}{Relative error with}\\
        & \multicolumn{4}{c}{CO/\cii{} template scale factors ...} \\\cline{2-5}
        Parameter & \replaced{Unused}{Fixed} & $A_\text{CO}$ \replaced{used}{free} & $A_\text{\cii{}}$ \replaced{used}{free} & Both \replaced{used}{free}\\\hline
$\Delta_T$  & 0.046\% & 0.089\% & 0.047\% & 0.09\% \\
$y$  & 0.41\% & 0.73\% & 0.42\% & 0.75\% \\
$\mu$  & 68\% & 160\% & 70\% & 160\% \\\hline
$A_D$  & 0.22\% & 0.25\% & 0.23\% & 0.26\% \\
$\beta_D$  & 0.062\% & 0.068\% & 0.064\% & 0.07\% \\
$T_D$  & 0.0055\% & 0.0056\% & 0.0055\% & 0.0056\% \\
$A_{\rm CIB}$  & 0.66\% & 0.77\% & 0.71\% & 0.82\% \\
$\beta_{\rm CIB}$  & 0.2\% & 0.22\% & 0.21\% & 0.23\% \\
$T_{\rm CIB}$  & 0.048\% & 0.055\% & 0.055\% & 0.063\% \\
$A_S$ & 6.5\% & 7.4\% & 6.5\% & 7.4\% \\
$\alpha_S$  & 9.2\% & 9.2\% & 9.4\% & 9.4\% \\
$\omega_S$  & 46\% & 47\% & 48\% & 48\% \\
$A_{FF}$ & 5.1\% & 7.5\% & 5.2\% & 7.7\% \\
$A_{\rm AME}$  & 1.8\% & 2.0\% & 1.8\% & 2.0\% \\\hline
$A_{\rm CO}$  &---& 6.4\% &---& 6.4\% \\
$A_{\rm [C\,II]}$  &---&---& 0.12\% & 0.12\% \\
    \end{tabular}\end{ruledtabular}
    \caption{Relative errors on spectral distortion and foreground parameters based on a Fisher forecast using \PIXIE sensitivities, with CO/\cii{} treated by way of uniformly scaling pre-calculated templates with $A_\text{CO}$ and $A_\text{\cii{}}$. We fix $kT_{\rm eSZ}$ and $\omega_2^{\rm eSZ}$ for this forecast. Note also the presence of \changeJfirstround{conservative} 10\% priors on $A_S$ and $\alpha_S$.}
    \label{tab:ACO_ACII}
\end{table}
We show the resulting relative errors on each parameter in~\autoref{tab:ACO_ACII}, in scenarios including none, one, or both of these scaling factors as free parameters. In this Fisher forecast, we expect that the sensitivity of \PIXIE is sufficient to constrain $A_\text{\cii{}}\neq0$ at $>800\sigma$, and $A_\text{CO}\neq0$ at $15\sigma$. Much of this constraining power depends on access to higher frequencies, particularly for the \cii{} emission. Cutting off the Fisher analysis at 600 GHz rather than 3 THz, we would expect to only constrain $A_\text{\cii{}}\neq0$ at $19\sigma$, and $A_\text{CO}\neq0$ at $9\sigma$.

Because of the highly distinct spectral structure and the brightness of our model CO and \cii{} spectra, the constraining power remains robust when we allow $kT_{\rm eSZ}$ and \replaced{$\omega_2^{\rm eSZ}$}{thus the relativistic Compton-$y$ distortion} to vary, as shown in our primary Fisher forecast of~\autoref{tab:CO_CII}. The detection significance for $A_\text{CO}\neq0$ and $A_\text{\cii{}}\neq0$ diminish only slightly from marginalising over the parameterisation of relativistic corrections to the $y$-type distortion.

\begin{table*}
    \centering
    \begin{ruledtabular}\begin{tabular}{rr@{\hspace{1ex}}lcccccccc}
          & \multicolumn{2}{c}{Fiducial value} &No CO/\cii{} & $A_\text{CO}$ & $\{A_\text{CO},A_\text{\cii{}}\}$ & \multicolumn{5}{c}{Four-parameter model for CO/\cii{}} \\\cline{7-11}
        Parameter & \multicolumn{2}{c}{(units)} & \PIXIE & \PIXIE & \PIXIE & CO LIM & \PIXIE & + CO LIM & \SPIXIE & + CO LIM\\\hline
$A_\text{CO}$ & 1& &---& 6.\replaced{7}{4}\% & 6.\replaced{7}{5}\% &---&---&---&---&---\\
$A_\text{\cii{}}$ & 1& &---&---& 0.1\replaced{4}{3}\% &---&---&---&---& ---\\\hline

$Z'_{\rm break}$ & 2 &&---&---&---& 65\% & 0.1\replaced{7}{6}\% & 0.16\% & 0.01\replaced{7}{6}\% & 0.016\% \\
$T_{k,2}$ & 0.5 & (K) &---&---&---& 34\% & 13\% & 10\% & 1.3\% & 1.3\% \\
$\gamma_{\rm MC}$ & 3& &---&---&---& 8.5\% & 0.84\% & 0.\replaced{4}{39}\% & 0.085\% & 0.07\replaced{7}{6}\% \\
$\delta_{\rm MC}$ & $2.2$&${}\times10^{-3}$ &---&---&---& 20\% & 1\replaced{1}{0}\% & 6.\replaced{7}{4}\% & 1.\replaced{2}{0}\% & \replaced{1.1}{0.99}\% \\\hline
$\Delta_T$ & $1.2$&${}\times10^{-4}$& 0.\replaced{12}{061}\% & 0.\replaced{15}{093}\% & 0.\replaced{16}{094}\% &---& 0.1\replaced{6}{1}\% & 0.\replaced{12}{08}\% & 0.0\replaced{62}{35}\% & 0.0\replaced{59}{35}\% \\
$y$ & $1.77$&${}\times10^{-6}$& \replaced{3.9}{1.1}\% & \replaced{4}{1}.2\% & \replaced{4.4}{1.2}\% &---& \replaced{4.6}{1.5}\% & \replaced{4}{1}.3\% & \replaced{1.2}{0.47}\% & \replaced{1.1}{0.46}\% \\
$kT_{\rm eSZ}$ & 1.282 &(keV) & \replaced{120}{19}\% &\replaced{120}{19}\% & \replaced{130}{21}\%  &---& \replaced{140}{34}\% & \replaced{130}{24}\% & \replaced{25}{6.3}\% & \replaced{25}{5.9}\% \\
\deleted{$\omega_2^{\rm eSZ}$ & 1.152 & & 760\% &780\% & 810\%  &---& 860\% & 840\% & 130\% & 120\% \\}
$\mu$ & $2$&${}\times10^{-7}$ & \replaced{160}{86}\% & \replaced{24}{16}0\% & \replaced{24}{16}0\% &---& \replaced{25}{19}0\% & 1\replaced{7}{1}0\% & \replaced{86}{49}\% & \replaced{80}{48}\% \\\hline
$A_D$ & 1.36 &(MJy\,sr$^{-1}$) & 0.\replaced{75}{28}\% & 0.\replaced{76}{32}\% & 0.\replaced{84}{36}\%  &---& 0.\replaced{96}{47}\% & 0.\replaced{90}{34}\% & 0.\replaced{10}{083}\% & 0.0\replaced{99}{82}\% \\
$\beta_D$ & 1.53 && 0.\replaced{22}{087}\% & 0.\replaced{22}{094}\% & 0.\replaced{25}{11}\% &---& 0.\replaced{28}{15}\% & 0.\replaced{27}{10}\% & 0.02\replaced{9}{1}\% & 0.02\replaced{9}{1}\% \\
$T_D$ & 21 &(K) & 0.0\replaced{21}{093}\% & 0.0\replaced{22}{096}\% & 0.0\replaced{24}{10}\% &---& 0.0\replaced{27}{15}\% & 0.0\replaced{25}{11}\% & 0.00\replaced{33}{16}\% & 0.00\replaced{33}{14}\% \\
$A_{\rm CIB}$ & 0.346 &(MJy\,sr$^{-1}$) & \replaced{2.0}{0.75}\%  & \replaced{2.0}{0.87}\% & \replaced{2.3}{1.0}\%  &---& \replaced{2.6}{1.2}\% & \replaced{2.5}{0.9}\% & 0.\replaced{3}{29}\% & 0.\replaced{3}{29}\% \\
$\beta_{\rm CIB}$ & 0.86 && 0.\replaced{65}{24}\% & 0.\replaced{65}{27}\% & 0.\replaced{73}{32}\% &---& 0.\replaced{83}{40}\% & 0.\replaced{7}{2}9\% & 0.0\replaced{88}{77}\% & 0.0\replaced{87}{76}\% \\
$T_{\rm CIB}$ & 18.8 &(K) & 0.07\replaced{6}{2}\% & 0.0\replaced{8}{75}\% & 0.0\replaced{8}{75}\% &---& 0.11\% & 0.0\replaced{87}{86}\% & 0.0\replaced{3}{2}8\% & 0.0\replaced{38}{27}\% \\
$A_S$ & 288 &(Jy\,sr$^{-1}$) & \replaced{7.2}{6.7}\% & \replaced{8.3}{7.5}\%  & \replaced{8.3}{7.5}\%  &---& \replaced{8.4}{7.7}\% & \replaced{7.3}{6.8}\% & 1.\replaced{6}{2}\% & \replaced{1.2}{0.99}\% \\
$\alpha_S$ & $-0.82$&& \replaced{10}{9.9}\% & \replaced{10}{9.9}\% & \replaced{10}{9.9}\% &---& 10\% & \replaced{10}{9.9}\% & \replaced{9.5}{7.2}\% & \replaced{9.5}{7.1}\% \\
$\omega_S$ & 0.2& & 5\replaced{2}{1}\%  & 5\replaced{2}{1}\% & 5\replaced{2}{1}\% &---& \replaced{52}{51}\% & \replaced{52}{51}\% & \replaced{48}{35}\% & \replaced{48}{35}\% \\
$A_{FF}$ & 300 &(Jy\,sr$^{-1}$) & \replaced{8}{6}.0\% & \replaced{10}{7.8}\% & \replaced{10}{7.9}\%  &---& \replaced{11}{9.2}\% & \replaced{8.4}{6.9}\% & \replaced{4.7}{2.9}\% & \replaced{4.5}{2.9}\% \\
$A_{\rm AME}$ & 1 & & \replaced{2.2}{1.8}\% & 2.\replaced{6}{0}\% & 2.\replaced{6}{0}\% &---& \replaced{3.1}{2.9}\% & 2.\replaced{7}{5}\% & 0.39\% & 0.3\replaced{7}{6}\%
    \end{tabular}\end{ruledtabular}
    \caption{Relative errors on spectral distortion and foreground parameters based on Fisher forecasts using the four-parameter model for the CO/\cii{} global signal, and some combination of CO LIM, \PIXIE, and/or \SPIXIE sensitivities. For comparison, we also show forecasts using only a uniform scaling of both CO and \cii{} templates as in~\autoref{tab:ACO_ACII} rather than the full model, although unlike in~\autoref{tab:ACO_ACII} we leave free the parameter\replaced{s $kT_{\rm eSZ}$ and $\omega_2^{\rm eSZ}$}{ $kT_{\rm eSZ}$ (but not $\omega_2^{\rm eSZ}$)} governing relativistic corrections to the $y$-type distortion. Note again the presence of 10\% priors on $A_S$ and $\alpha_S$.}
    \label{tab:CO_CII}
\end{table*}

\autoref{tab:CO_CII} also shows the fundamental sensitivity of \PIXIE and \SPIXIE to the CO/\cii{} global signal in terms of our four-parameter model, which we also represent as Fisher forecast confidence ellipses for the four CO/\cii{} parameters in~\cref{fig:ellipses}. With \PIXIE data, the degeneracies between the four parameters are mostly mild, except for between the overall CO intensity normalisation controlled by $\delta_\text{MC}$ and the redshift evolution of the kinetic temperature controlled by $T_{k,2}$. We anticipated this in~\autoref{sec:fishermethods} based on an examination of the derivatives of the spectrum with respect to those two parameters: increasing only the intensity of the higher-$J$ CO lines results in a modest skew at best relative to increasing the intensity of CO emission uniformly. Nonetheless, as we also anticipated, the two are not entirely degenerate, and we expect the \PIXIE (\SPIXIE) concept to have the fundamental sensitivity to achieve $\approx10\%$ ($\approx1\%$) constraints on both variables. A \PIXIE-like (\SPIXIE) mission would also constrain the remaining parameters at sub-percent (sub-per-mille) levels, expected given their relative lack of degeneracy and the high significance forecast for constraining the overall CO/\cii{} amplitudes $A_\text{CO}$ and $A_\text{\cii{}}$ in the simpler case.
\begin{figure}
    \centering
    \includegraphics[width=0.96\linewidth]{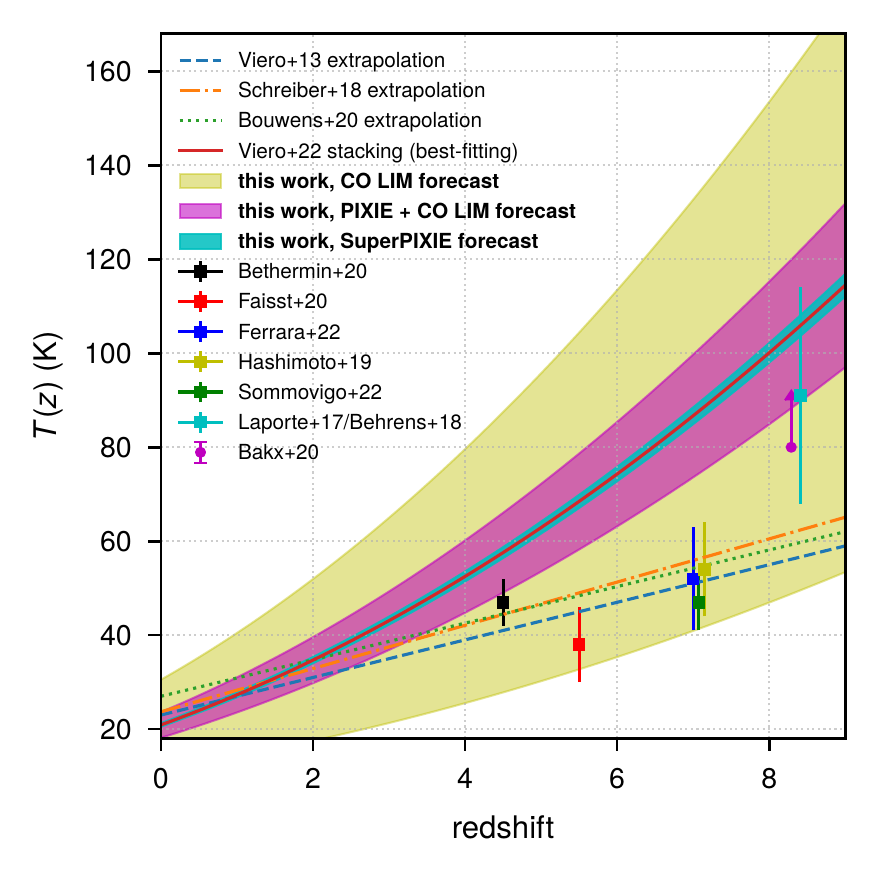}
    \caption{Evolution of dust temperature with redshift seen through measurements of individual objects~\citep{Bethermin20,Faisst20,Ferrara22,Hashimoto19,Sommovigo22,Laporte17,Behrens18,Bakx20}, extrapolations from findings at lower redshift~\citep{Viero13,Schreiber18,Bouwens20}, or stacking~\citep{Viero22}; the last result directly provides the kinetic temperature evolution for our fiducial extragalactic CO/\cii{} model. For comparison, we project as $T(z)$ the expected constraints on $T_{k,2}$ given CO LIM, \PIXIE, and \SPIXIE sensitivities.}
    \label{fig:Tz}
\end{figure}

The CO LIM measurements have moderate constraining power on each of the four parameters, but given their focus in this specific case on two specific cosmic snapshots ($z\sim3$ and $z\sim7$), we find a somewhat limited ability to constrain the picture of molecular gas across all redshifts. Nonetheless, as should be clear from the improvement in constraints on $\mu$ and other parameters, they do provide real constraining power on the CO/\cii{} model. The primary exception is $Z'_\text{break}$, likely because at the redshifts surveyed by CO LIM, the host halos dominating the signal lie below the break mass of $M_h=10^{12.3}\,M_\odot$. For the other parameters, combining \PIXIE with CO LIM data---without even directly cross-correlating the two data sets---would provide 30--100\% improvements in sensitivity over \PIXIE alone, although the relative improvement with CO LIM is far more modest at face value when considering \SPIXIE.

We can graphically illustrate the meaning of the constraint on $T_{k,2}$ by plotting the corresponding interval of $T_k(z)$, as shown in~\cref{fig:Tz}. The work of~\cite{Viero22} suggests the possibility that dust temperatures at the late-reionisation epoch are much higher than suggested by either extrapolations from lower redshift (e.g.,~\citep{Viero13,Schreiber18,Bouwens20}) or analyses of individual objects (e.g.,~\citep{Bethermin20,Faisst20,Ferrara22,Hashimoto19,Sommovigo22,Laporte17,Behrens18,Bakx20}). The CO kinetic temperature in turn can act as a proxy for dust temperature, owing to the tendency for CO and dust to be co-located in dense regions and even interact in key ways that govern the cooling and chemistry of molecular clouds (cf., e.g.,~\citep{Hocuk16}). We find that the constraint on $T_{k,2}$, interpreted in the most na\"{i}ve possible way as a direct proxy for dust temperature evolution, will be competitive with constraints on individual objects with \PIXIE sensitivities. The added constraining power with \SPIXIE will actively allow adjudication on whether the CO-emitting population follows dust temperatures similar to or elevated relative to individual detection.

In reality, the interpretation of kinetic temperature in relation to dust is complex and depends on molecular cloud parameters like H$_2$ density that are held fixed in our model. Probing these will require sensitivity to further degrees of freedom, and is likely an area of research where CO LIM experiments and absolute spectrometers will have a synergistic effect through a coordinated joint analysis that includes direct cross-correlation, combining low-resolution absolute calibration with mid-resolution anisotropy cubes sensitive to fluctuations and tracer bias.

\begin{figure}
    \centering
    \includegraphics[width=0.96\linewidth]{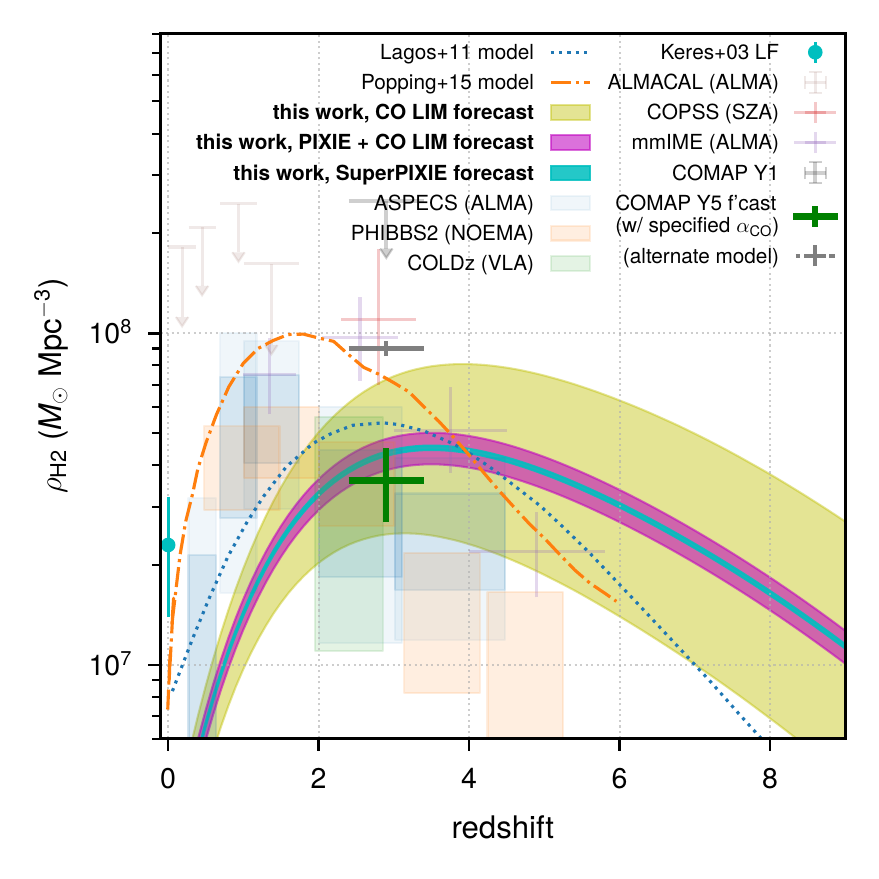}
    \caption{Same as the lower panel of~\cref{fig:COcheck}, but with forecasts both for near-term COMAP sensitivities (given two different CO models at $z\sim3$), and for the experiments considered in this work. Note that the COMAP forecasts~\citep{COMAPESV} fix the conversion between CO and H$_2$, whereas we directly infer this conversion via the modelling in this work, meaning the resulting constraints should not be considered in the same fashion.}
    \label{fig:rhoH2}
\end{figure}

We also show the constraints on $\gamma_\text{MC}$ and $\delta_\text{MC}$ projected into the space of cosmic molecular gas mass density as a function of redshift, in~\cref{fig:rhoH2}. The result is once again competitive with surveys of resolved objects and even forecasts of the constraining power of single-dish LIM surveys like the COMAP Pathfinder~\citep{COMAPESV} that assume fixed conversion factors between CO and H$_2$. However, here also, we have held parameters like average cloud mass fixed in our modelling of the global signal, and finer modelling of CO excitation conditions will likely demand the direct cross-correlation of monopole and anisotropy data.

\subsection{Carbonic lines as foregrounds: comparing uncertainties in relic distortion parameters}
\label{sec:res_fg}

Recall that in the simplest case, we forecast the effect of the CO and \cii{} foreground emission on the error for other parameters by introducing fixed CO and \cii{} templates with scaling factors for each. We return first to~\autoref{tab:ACO_ACII} to review the effect of accounting for carbonic lines on our parameters.

The most drastic effect is on $\mu$, the error on which increases from $\sigma_\mu=1.4\times10^{-7}$ to $\sigma_\mu=3.1\times10^{-7}$---although we note that even in the latter case, the corresponding $2\sigma$ limit of $|\mu|<6.3\times10^{-7}$ would be almost a 75-fold improvement over the current best limit from~\cite{BianchiniFabbian22}. With the relativistic correction parameters fixed, the effect on $\sigma_y$ is also significant although not catastrophic, with $y/\sigma_y$ decreasing from 240 to 130. \replaced{When we allow}{Allowing} the relativistic correction parameters to vary\added{, however, has a much greater effect on $\sigma_y$} for our primary forecasts in~\autoref{tab:CO_CII}\added{, taking $y/\sigma_y$ from 240 to 90. In comparison}, the effect on $\sigma_y$ from introducing $A_\text{CO}$ and $A_\text{\cii{}}$ becomes modest relative to the effect from \added{just }introducing $kT_{\rm eSZ}$ \replaced{and $\omega_2^{\rm eSZ}$}{as an additional free parameter}, with the detection significance for $y$ reduced from \replaced{$\sim25\sigma$ to $\sim23\sigma$}{$\sim90\sigma$ to $\sim83\sigma$} due to $A_\text{CO}$ and $A_\text{\cii{}}$. The $2\sigma$ limit on the $\mu$-type distortion \replaced{also degrades slightly to $|\mu|<9.6\times10^{-7}$, still almost a 50}{degrades to a baseline of $|\mu|<3.4\times10^{-7}$ with relativistic Compton-$y$ corrections allowed to vary (versus $|\mu|<2.8\times10^{-7}$ without). However, in the presence of carbonic line models the constraint does not further degrade with versus without allowing $kT_{\rm eSZ}$ to vary, maintaining the near-75}-fold improvement over the limit from~\cite{BianchiniFabbian22}.

\replaced{Interestingly}{Between introducing $A_\text{CO}$ and introducing $A_\text{\cii{}}$}, a majority of the penalty in error is introduced with $A_\text{CO}$, in both CMB and foreground spectral features. The primary exceptions are in the CIB and Galactic dust parameters, likely due to our specific fiducial form for both modified blackbody spectra peaking within the frequency range spanned by the \cii{} spectrum. Meanwhile, the error on the synchrotron component is entirely unaffected, which is sensible given that the prior on $\alpha_S$ likely provides most of the information. \changeJfirstround{However, one should note that in the future priors on synchrotron foregrounds at the percent level or better can be anticipated. In addition, development of on-chip FTS technology \citep[e.g.,][]{Thakur2023} for measuring CMB spectral distortions at low frequencies may provide significant gains, overall rendering our estimates quite conservative.}

Introducing more parameters to control the CO and \cii{} components has a fairly modest effect on parameter uncertainties. As~\autoref{tab:CO_CII} shows, when we use the four-parameter model instead of fixed templates, we do see a modest increase in the uncertainty on $A_\text{AME}$, likely because the lower-frequency part of the CO spectrum may now vary differently from the higher-frequency part. Other parameters also see some penalty, in particular CIB and Galactic dust parameters once again. \replaced{However}{For the CMB spectral distortion parameters}, the increase in degrees of freedom has \replaced{minimal}{a $\approx20$--25\% } effect on uncertainties for $\mu$\replaced{---or indeed}{ and} $y$, \replaced{whose}{and the latter's} detection significance falls \replaced{only slightly further to $\sim22\sigma$ with}{further from $\sim83\sigma$ to $\sim67\sigma$ with} \PIXIE alone.

\begin{figure}
    \centering
    \includegraphics[width=0.9486\linewidth]{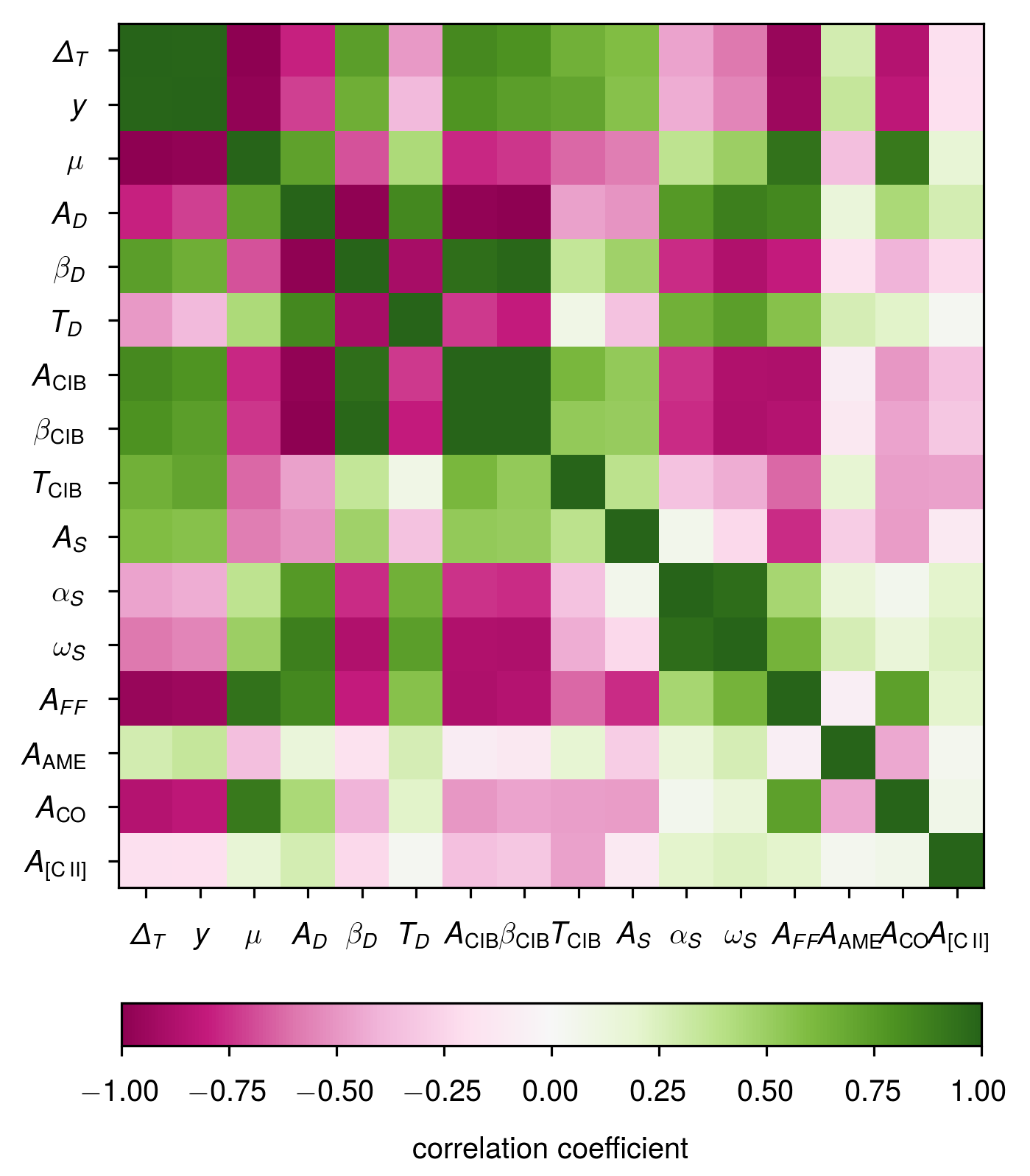}
    \includegraphics[width=0.9486\linewidth]{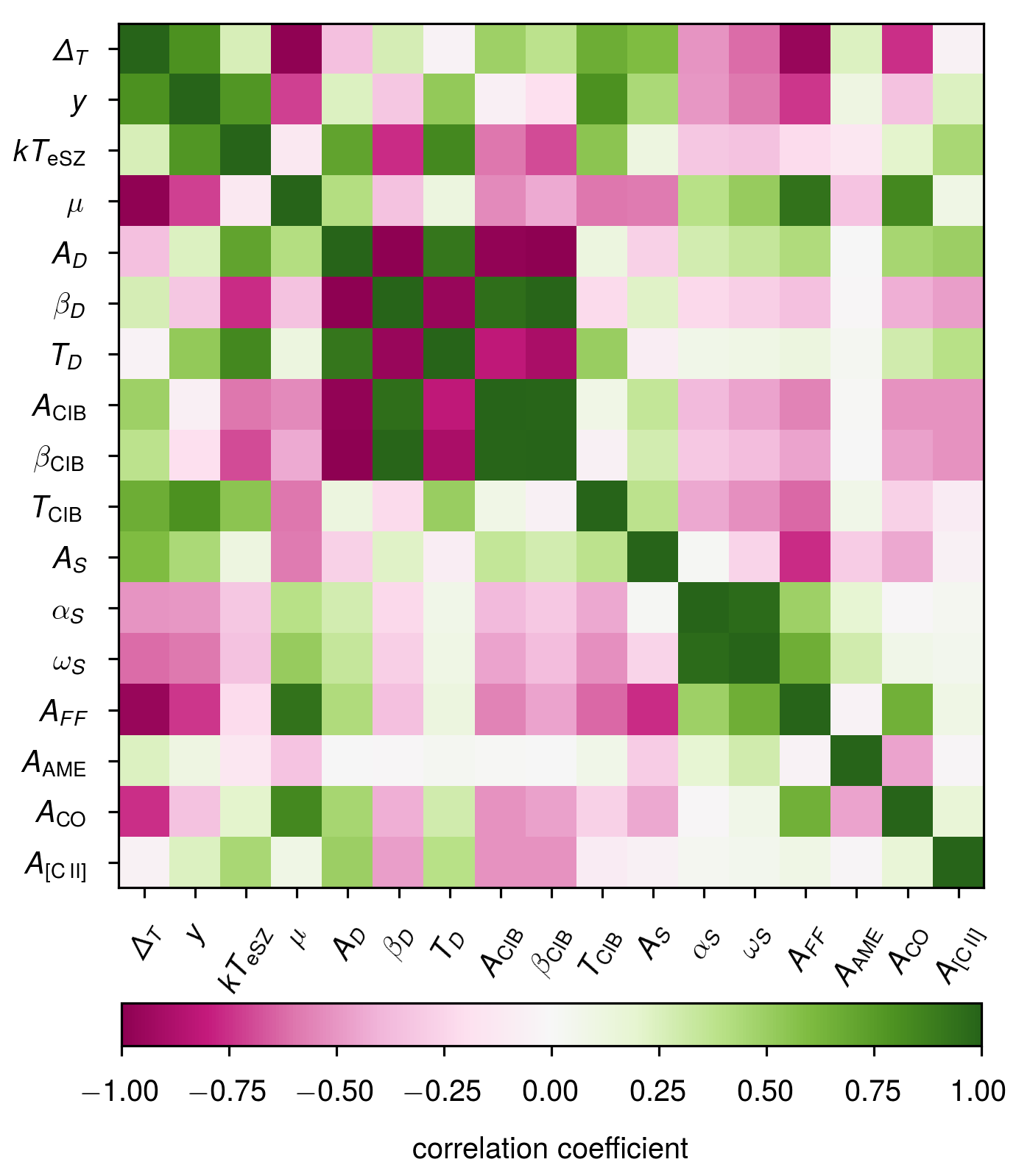}
    \caption{Inter-parameter correlation coefficients derived from the Fisher analysis in the main text, with $kT_{\rm eSZ}$ and $\omega_2^{\rm eSZ}$ fixed (upper panel) and then \added{$kT_{\rm eSZ}$ }allowed to vary (lower panel).}
    \label{fig:cov}
\end{figure}

One factor in the impact on $y$ in particular is a natural confluence of important CO and \cii{} emission frequencies with the SZ cross-over frequency of $\approx218$\,GHz. This frequency corresponds to a \cii{} redshift of 7.7, where the \cii{} emission is on a clear downward trend with decreasing frequency, due to lower gas-phase metallicities and fewer objects exceeding the atomic cooling limit at higher redshift. Meanwhile, the cross-over frequency also corresponds to CO(2--1), CO(3--2), and CO(4--3) redshifts of 0.05, 0.6 and 1.1---all redshifts where CO emission will clearly get brighter with increasing redshift in approaching the peak of cosmic molecular gas content and star-formation activity. This results in a mild resemblance between a $y$-type distortion and the combination of dimmer-than-expected CO with additional degrees of freedom provided by \cii{} and other foreground components around the cross-over frequency, which is not overly dependent on more degrees of freedom to control the shapes of the CO and \cii{} sky-averaged spectra. 

Since the CO/\cii{} global signal does not exist in a vacuum, their effect on $\sigma_y$ depends on our modelling of other effects. \replaced{For example, when fixing relativistic corrections to the $y$-type distortion, most of the effect on constraining power arises when we introduce $A_\text{CO}$. On the other hand, when varying $kT_{\rm eSZ}$ and $\omega_2^{\rm eSZ}$, the effect on constraining power on $y$ arises more or less in equal parts with the introduction of $A_\text{CO}$ and the introduction of $A_\text{\cii{}}$ (though we do not show this in~\autoref{tab:CO_CII})---but only because $y$ becomes more strongly correlated with $kT_{\rm eSZ}$ and $\omega_2^{\rm eSZ}$, as opposed to any increased relative importance ascribable to $A_\text{\cii{}}$.}{Allowing relativistic corrections to the $y$-type distortion to vary, for instance, introduces additional parameter covariances affecting the covariance between $A_\text{CO}$ and $y$, and thus the size of the effect on $\sigma_y$ of varying $A_\text{CO}$.} This is clear when the parameter covariance matrix is used to conduct a correlation analysis, as graphically illustrated in~\cref{fig:cov}\added{, which also clearly shows significant covariance between $y$ and other foreground parameters}. In any case, only a comparatively small additional effect arises from introducing the four-parameter model over $A_\text{CO}$ and $A_\text{\cii{}}$.

The overall impact on the detection significance for $y$ ends up being far from catastrophic. The primary reason for this is that the magnitude of the $y$-distortion is several times larger than the expected deviation due to extragalactic CO/\cii{} emission in each one of many frequency channels that \PIXIE or \SPIXIE would measure across the microwave spectrum. Furthermore, information at high frequencies will likely be sufficient to distinguish the signature of extragalactic \cii{} from a $y$-type distortion. The latter essentially vanishes beyond 800 GHz; by contrast, the \cii{} global signal will likely rise almost monotonically from the $z=0$ emission frequency of 1900.5 GHz all the way down to $\approx600$ GHz corresponding to $z\sim2$ during the so-called `cosmic noon'. As a result, a small level of variation in $y$ requires very large shifts in the CO/\cii{} global signal combined with shifts in numerous uncorrelated foregrounds.

The reason for the \replaced{larger}{similar} impact on $\mu$ compared to $y$ is not obvious at first, given that the $\mu$-type distortion has a lower cross-over frequency at 125 GHz. However, as with the $y$-type distortion, it is not the CO foreground \emph{per se}, but the potential combination of the CO foreground---which peaks near the cross-over frequency for the $\mu$-type distortion---with deviations due to other components like $\Delta_T$, spinning dust, and so on. Foreground combinations can quite closely mimic a $\mu$-type distortion without reaching the level of contrivance needed to mimic a $y$-type distortion, resulting in the relative doubling in error seen in~\autoref{tab:ACO_ACII} simply by introducing $A_\text{CO}$ as an additional nuisance parameter, and \replaced{the 50\%}{a similar} increase in error in~\autoref{tab:CO_CII}.

\begin{figure}
    \centering
    \includegraphics[width=0.96\linewidth]{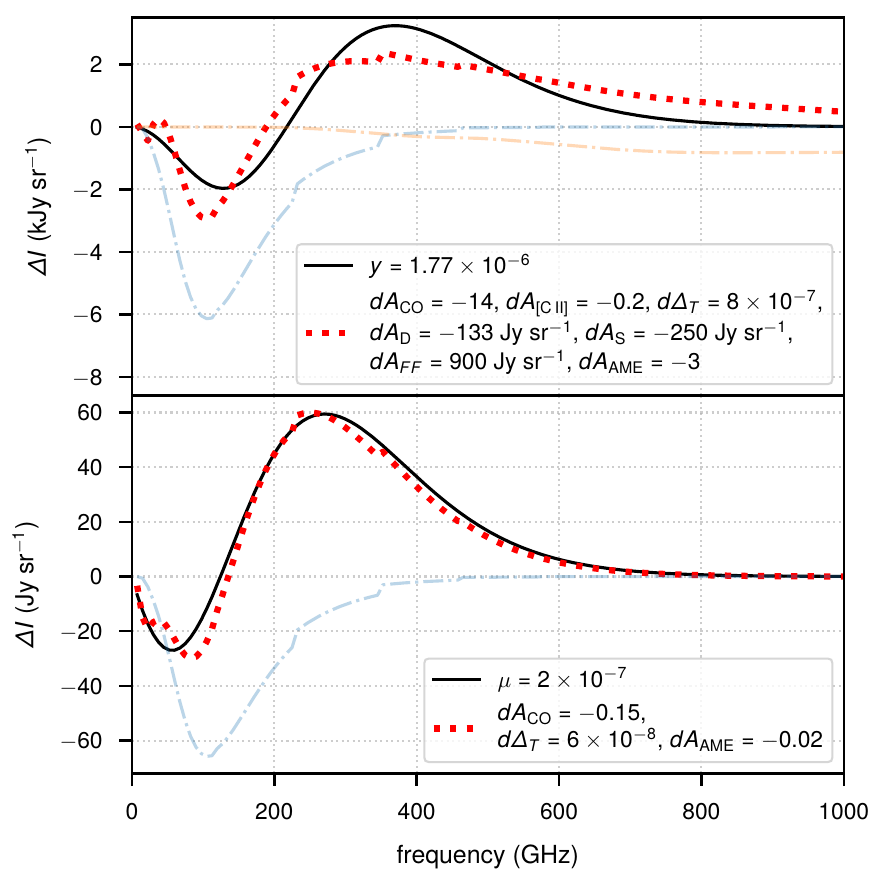}
    \caption{Fiducial $y$- and $\mu$-type distortions (upper and lower panels respectively) compared to possible changes in the spectrum due to shifts in foreground parameters (as indicated in the legends), including scaling of CO and \cii{} templates (shown separately in faint blue and orange dash-dotted curves). Note the \emph{thousandfold} difference in vertical scale between the upper panel (in kJy\,sr$^{-1}$) and the lower panel (in Jy\,sr$^{-1}$). Recall that the mission sensitivity expected for \PIXIE (\SPIXIE) is 5.1 (0.5) Jy\,sr$^{-1}$.}
    \label{fig:pathology}
\end{figure}

\autoref{fig:pathology} more explicitly shows examples of extragalactic CO/\cii{} combined with other foreground components to mimic $\mu$- and $y$-type distortions. As previously discussed, the signature of CO/\cii{} emission clearly differs from a $y$-type distortion at higher frequencies, by many times the mission sensitivity (noting that the upper panel of~\cref{fig:pathology} shows the change to the monopole signal in units of kJy\,sr$^{-1}$, not Jy\,sr$^{-1}$). However, the combination of CO and other foreground components can mimic a $\mu$-type distortion within measurement uncertainties with only $\lesssim10\%$ deviations in parameter values, recalling that our fiducial value for $\mu$ is an \changeJfirstround{enhanced} one at ten times the standard cosmological expectation. The specific combinations we show in~\cref{fig:pathology} are unlikely to manifest by chance, but marginalising over such cases is what leads to the 10\% and 50\% relative increases in uncertainty in $y$ and $\mu$ respectively when introducing our CO/\cii{} model, as shown above in~\autoref{tab:CO_CII}.

Nonetheless, as already stated, the CO and \cii{} foregrounds are far from presenting a significant roadblock to a confident detection of $y$-type observables. In fact, if we fold in CO LIM measurements, we actually find that \PIXIE with CO LIM information returns the constraining power on $y$ roughly to the level achieved using fixed templates rather than the four-parameter model. The addition of CO LIM information also \added{largely }returns the constraining power on $\mu$ \replaced{roughly to}{towards} the level achieved without marginalising over CO/\cii{} foreground properties at all, from \replaced{$\sigma_\mu=5\times10^{-7}$ to $\sigma_\mu=3.4\times10^{-7}$, or from $|\mu|<10^{-6}$ to $|\mu|<6.8\times10^{-7}$ (95\% CL). While far from sufficient to measure our fiducial value of $\mu=2\times10^{-8}$, such sensitivity to $\mu$ would finally meet the goal set out by~\cite{FIRASFinal} for a `FIRAS II' to set limits on $\mu$ at the $\sim10^{-7}$ level}{$\sigma_\mu=3.9\times10^{-7}$ to $\sigma_\mu=2.3\times10^{-7}$, or from $|\mu|<7.7\times10^{-7}$ to $|\mu|<4.6\times10^{-7}$ (95\% CL)}. The subsequent \SPIXIE scenario then allows for a significantly improved $\approx80\sigma$ detection of $y\neq0$ without any priors imposed on the four CO and \cii{} parameters, improving moderately to $\approx90\sigma$ with CO LIM information. With the combination of \SPIXIE and CO LIM data, we also forecast \replaced{$\sigma_\mu=1.6\times10^{-7}$}{$\sigma_\mu=9.7\times10^{-8}$, just sufficient for a tentative detection of our fiducial $\mu$ value. Such sensitivity to $\mu$ would finally meet the goal set out by~\cite{FIRASFinal} for a `FIRAS II' to set limits on $\mu$ at the $\sim10^{-7}$ level}.

\subsection{Discussion: dependence of results on line emission modelling}

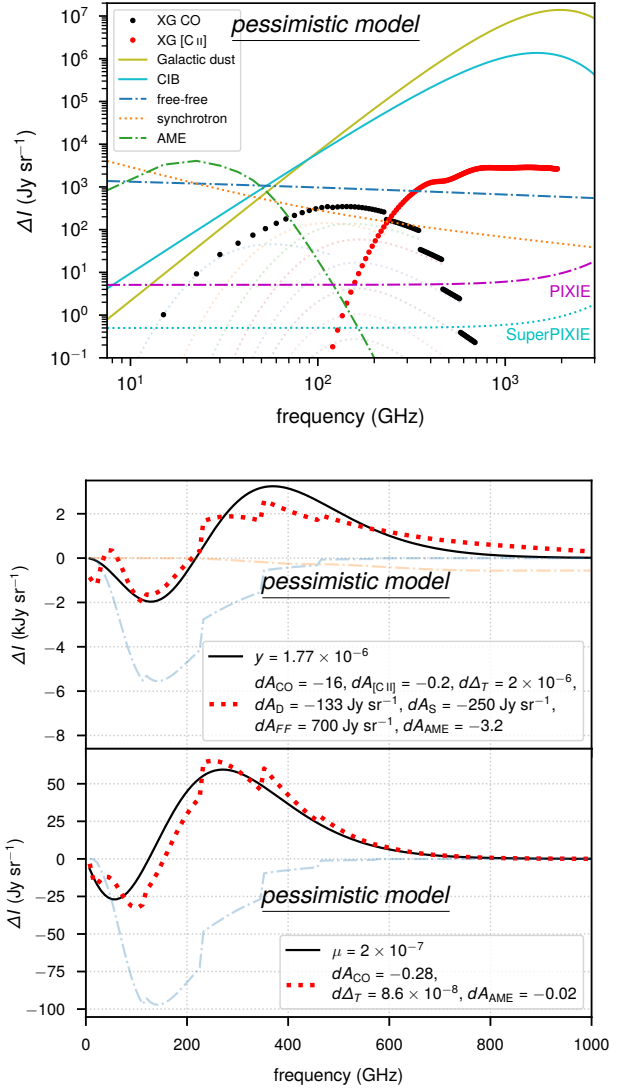
\begin{figure}
    \centering
    \begin{tikzpicture}
    \node () at (0,0) {\includegraphics[width=0.96\linewidth]{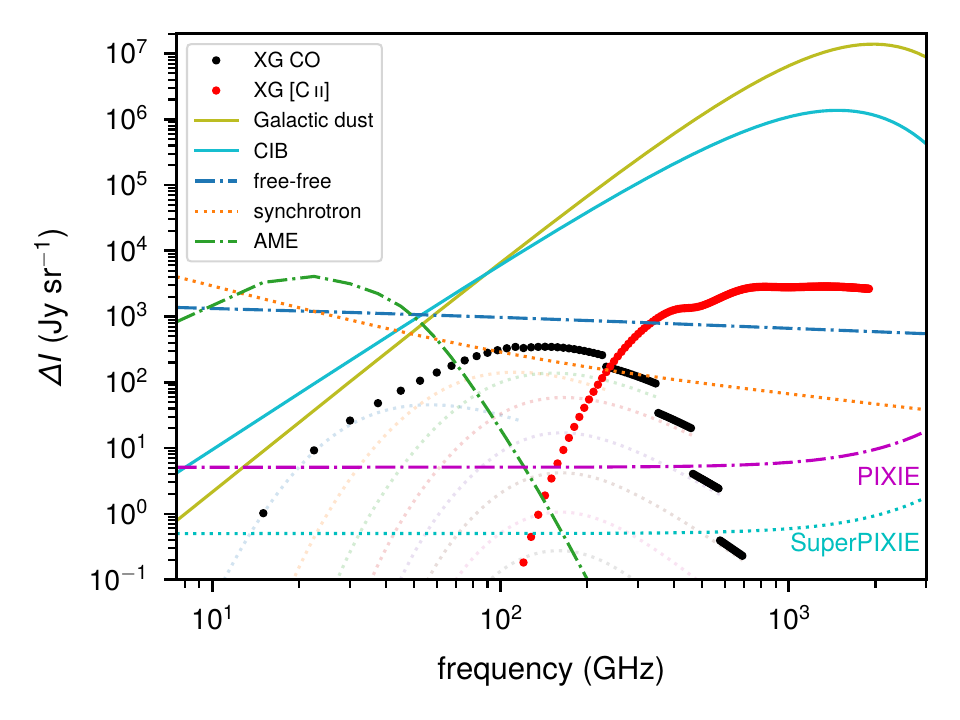}};
    \node () at (0.286,2.486) {\it\fontfamily{qhv}\selectfont \underline{pessimistic model}};
    \end{tikzpicture}
    
    \begin{tikzpicture}
    \node () at (0,0) {\includegraphics[width=0.96\linewidth]{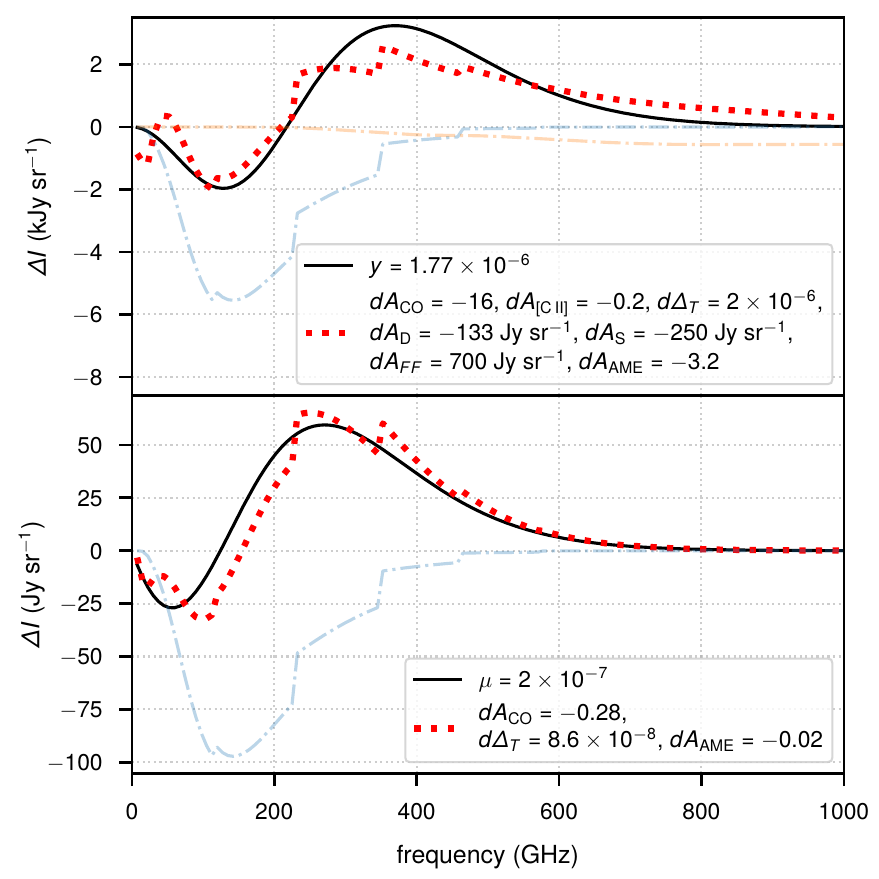}};
    \node () at (0.686,2.586) {\it\fontfamily{qhv}\selectfont \underline{pessimistic model}};
    \node () at (0.686,-1.586) {\it\fontfamily{qhv}\selectfont \underline{pessimistic model}};
    \end{tikzpicture}
    \caption{Same as the left-hand panel of~\cref{fig:breakdown} (upper panel) and same as~\cref{fig:pathology} (lower panels), but for a more pessimistic variation on our CO/\cii{} model with $\gamma_\text{MC}=1.86$ and $\delta_\text{MC}=5.5\times10^{-4}$.}
    \label{fig:pess}
\end{figure}
\begin{figure}
    \centering
    \begin{tikzpicture}
    \node () at (0,0) {\includegraphics[width=0.96\linewidth]{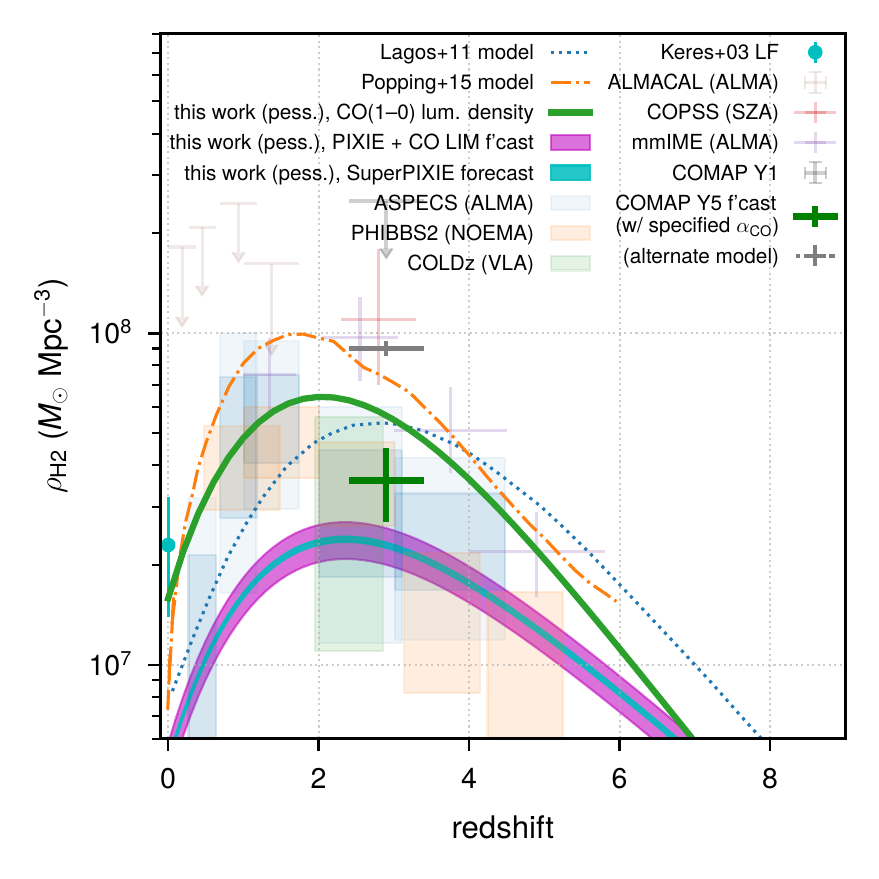}};
    \node () at (1.86,0.86) {\it\fontfamily{qhv}\selectfont \underline{(pess.): pessimistic model}};
    \end{tikzpicture}
    \caption{Same as~\cref{fig:rhoH2}, but for a more pessimistic variation on our CO/\cii{} model with $\gamma_\text{MC}=1.86$ and $\delta_\text{MC}=5.5\times10^{-4}$. In place of the CO LIM forecast, which does not strongly constrain the evolution of $\rho_\text{H2}$, we show the CO(1--0) luminosity density predicted by this pessimistic model variation, in the same way as in the lower panel of~\cref{fig:COcheck}.}
    \label{fig:rhoH2_pess}
\end{figure}

\begin{table*}
    \begin{ruledtabular}\begin{tabular}{cccccccc}
    & No CO/\cii{} & \multicolumn{3}{c}{Fixed CO/\cii{} templates, \PIXIE} & \multicolumn{3}{c}{Four-parameter model for CO/\cii{}}\\\cline{3-5}\cline{6-8}
        Parameter & \replaced{models}{\PIXIE} & CO only & \cii{} only & CO/\cii{} & \PIXIE & \PIXIE + CO LIM & \SPIXIE
    \\\hline
$\Delta_T$  & 0.\replaced{12}{061}\% & 0.\replaced{12}{070}\% & 0.\replaced{12}{061}\% & 0.\replaced{13}{070}\% & 0.\replaced{13}{075}\% & 0.\replaced{1}{07}2\% & 0.0\replaced{56}{37}\% \\
$y$  & \replaced{3.9}{1.1}\% & \replaced{3.9}{1.1}\% & \replaced{4.2}{1.1}\% & \replaced{4.2}{1.1}\% & \replaced{4.3}{1.8}\% & \replaced{4.3}{1.7}\% & \replaced{1.1}{0.52}\% \\
$kT_{\rm eSZ}$  & \replaced{120}{19}\% & \replaced{120}{19}\% & \replaced{13}{2}0\% & \replaced{13}{2}0\% & \replaced{130}{39}\% & \replaced{130}{35}\% & \replaced{2}{7.}2\% \\
\deleted{$\omega_2^{\rm eSZ}$  & 760\% & 760\% & 790\% & 790\% & 830\% & 820\% & 110\% \\}
$\mu$  & \replaced{160}{86}\% & 1\replaced{7}{0}0\% & \replaced{170}{86}\% & 1\replaced{7}{0}0\% & 1\replaced{8}{1}0\% & \replaced{170}{98}\% & \replaced{77}{50}\%\\\hline
$A_{\rm CO}$  &---& 4.3\% &---& 4.3\% &---&---& --\\
$A_{\rm [C\,II]}$  &---&---& 0.1\replaced{4}{3}\% & 0.1\replaced{4}{3}\% &---&---& --\\\hline
$Z'_{\rm break}$ &---&---&---&---& 0.16\% & 0.16\% & 0.016\% \\
$T_{k,2}$ &---&---&---&---& 12\% & 11\% & 1.2\% \\
$\gamma_{\rm MC}$ &---&---&---&---& 1.\replaced{1}{0}\% & 0.8\replaced{9}{7}\% & 0.11\% \\
$\delta_{\rm MC}$ &---&---&---&---& 11\% & 8.4\% & 1.1\% \\
    \end{tabular}\end{ruledtabular}
    \caption{Fisher forecasts for relative error on CMB spectral distortion and CO/\cii{} parameters given a more pessimistic variation on our CO/\cii{} model with $\gamma_\text{MC}=1.86$ and $\delta_\text{MC}=5.5\times10^{-4}$.}
    \label{tab:pess}
\end{table*}
One may question whether the results we have shown for CO/\cii{}, either as foreground or as signal, will change significantly based on the brightness of CO and \cii{} and its redshift evolution. We have already shown that the size of the effect of CO and \cii{} on $\sigma_y$ and $\sigma_\mu$ depends even on the parameterisation of other sources of spectral deviations. As such, the magnitude of the CO/\cii{} global signal will also inevitably change the magnitude of the effects we have predicted, but we can show that the qualitative results remain the same.

To establish this, we make a very basic adjustment to our four-parameter model, changing the values of $\gamma_\text{MC}$ and $\delta_\text{MC}$ from their fiducial values of 3 and $2.2\times10^{-4}$ to more pessimistic values of 1.86 and $5.5\times10^{-4}$. This is pessimistic in two senses:
\begin{itemize}
    \item The contribution from emission at $z\sim0$ is much larger than in the fiducial case, accentuating the `staircase' shape to the CO spectrum and making it more distinct from the smooth relic spectral distortions.
    \item The contribution from high redshift, on the other hand, decreases significantly, reducing the information gain from CO LIM.
\end{itemize}
We present some predictions of this more pessimistic variation in~\cref{fig:pess}, both in terms of the overall CO/\cii{} global signal and in terms of how they can still mimic relic spectral distortions. Note, however, that this mimicry now requires changes in foreground model parameters larger by factors of several than in the case of~\cref{fig:pathology} with the fiducial CO/\cii{} model variation. This in turn explains the parameter errors recalculated in~\autoref{tab:pess} for the pessimistic CO/\cii{} model, which shows increases in error dulled by factors of several compared to our fiducial results in~\autoref{tab:ACO_ACII} and~\autoref{tab:CO_CII}. The greatest effect is still on $\mu$, the error on which takes a $\sim10\%$ hit.

Interestingly, the constraining power of \PIXIE and \SPIXIE on our four-parameter model sees no significant penalty relative to the fiducial case, with the relative error on all parameters essentially the same. The information content contributed by CO LIM, on the other hand, decreases significantly. Under this pessimistic model variation, the total detection significance across all CO LIM observables falls to 42, but in particular the signal-to-noise ratio decreases by an order of magnitude for the $z\sim7$ observables (the auto and cross power spectra involving the 15 GHz observations), with the detection significance cut only in half for the 30 GHz auto spectrum (a mix of CO emission from $z\sim3$ and $z\sim7$). The broad frequency coverage of our simulated CMB spectral distortion experiments means that they, unlike CO LIM, do not have to rely on the brightness of the CO/\cii{} signal in survey volumes at specific redshift intervals. This likely allows \PIXIE and \SPIXIE to be more robust to this particular model variation, maintaining constraining power on the cosmic molecular gas density history as shown in~\cref{fig:rhoH2_pess}.

\added{
\begin{figure}
    \centering
    \includegraphics[width=0.986\linewidth]{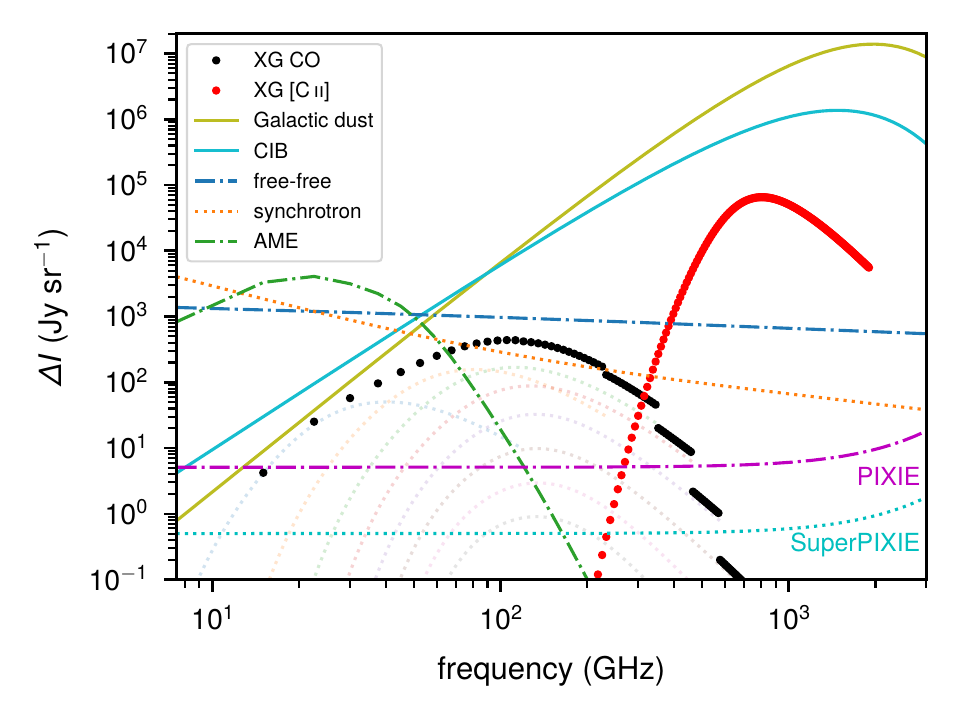}
    
    \includegraphics[width=0.986\linewidth]{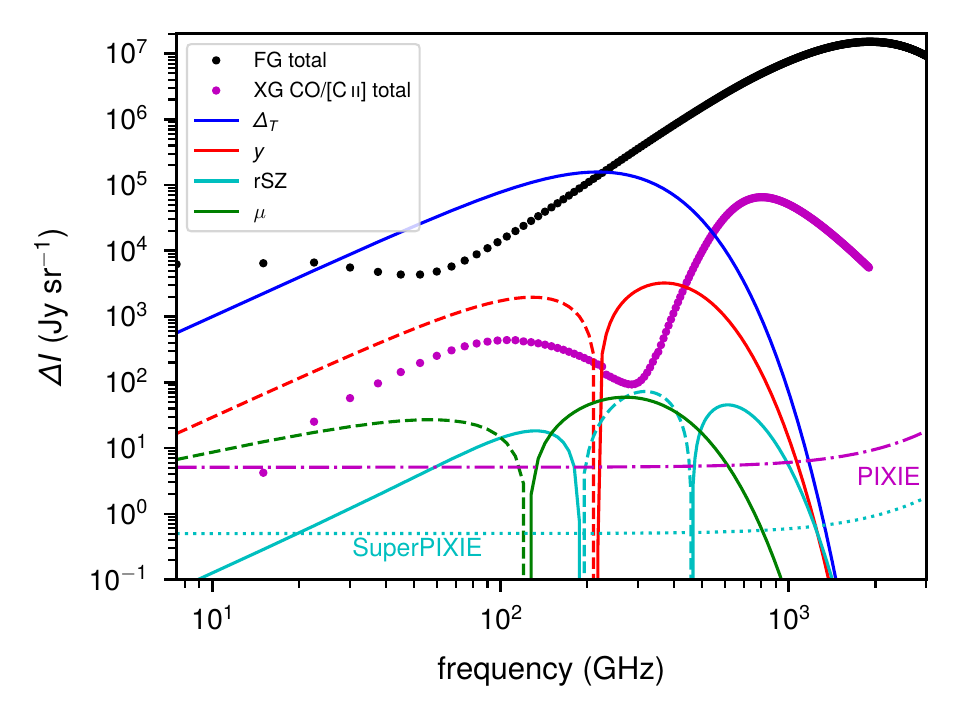}
    \caption{Sams as~\cref{fig:breakdown}, but using the fiducial CO model with a \cii{} model based on~\cite{Padmanabhan19}.}
    \label{fig:breakdown_pad18}
\end{figure}

\begin{table}[]
    \centering
    \begin{ruledtabular}\begin{tabular}{rcccc}
        & \multicolumn{4}{c}{Relative error with}\\
        & \multicolumn{4}{c}{CO/\cii{} template scale factors ...} \\\cline{2-5}
        Parameter & Fixed & $A_\text{CO}$ free & $A_\text{\cii{}}$ free & Both free\\\hline
$\Delta_T$  & 0.061\% & 0.093\% & 0.087\% & 0.1\% \\
$y$  & 1.1\% & 1.2\% & 2.5\% & 2.5\% \\
$kT_{\rm eSZ}$  & 19\% & 19\% & 55\% & 58\% \\
$\mu$  & 86\% & 160\% & 120\% & 170\% \\\hline
$A_{\rm CO}$  & --- & 6.4\% & --- & 6.7\% \\
$A_{\rm [C\,II]}$  & --- & --- & 0.038\% & 0.04\% \\
    \end{tabular}\end{ruledtabular}
    \caption{Same as the first four columns of~\cref{tab:pess}\addedtwo{ (but compare also the first three columns of~\cref{tab:CO_CII})}, but using a combination of our fiducial CO model with a \cii{} model based on~\cite{Padmanabhan19}.}
    \label{tab:pad18}
\end{table}

\begin{figure}
    \centering
    \includegraphics[width=0.9486\linewidth]{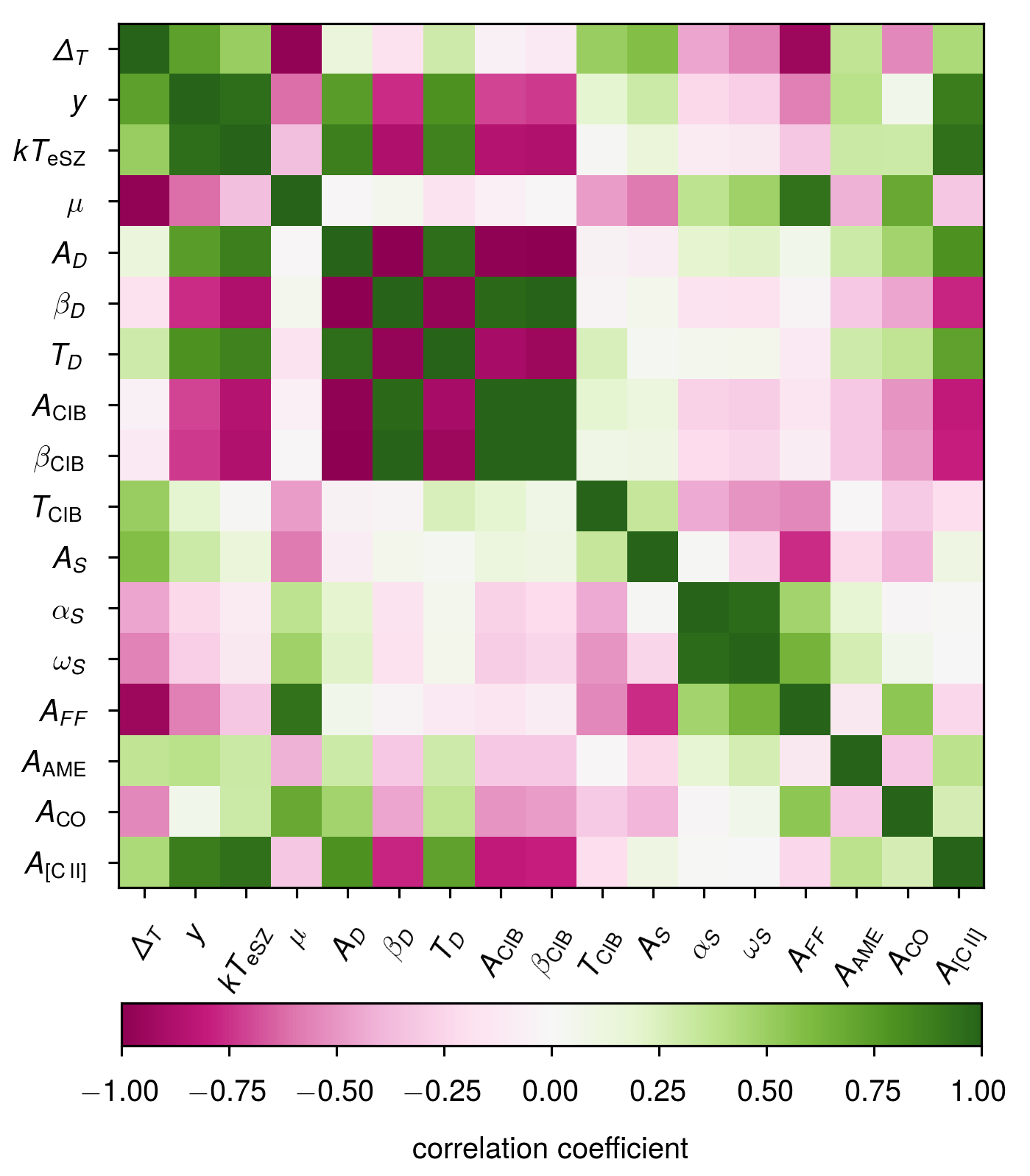}
    \caption{Inter-parameter correlation coefficients as in the lower panel of~\cref{fig:cov}, but from the Fisher analysis using the \cii{} model of~\cite{Padmanabhan19}.}
    \label{fig:cov_pad18}
\end{figure}}

\added{We can also consider variations of the \cii{} model, changing the spectral template for global \cii{} emission from our fiducial model to one based on~\cite{Padmanabhan19}. As~\cref{fig:CIIcheck} shows, the global average \cii{} line intensity of~\cite{Padmanabhan19} peaks more sharply at $z\sim2$ and declines more rapidly at higher redshift than our fiducial model. The consequent na\"{i}ve expectation would be that a \cii{} template of this form should distinguish itself more sharply from the primordial CMB distortions; however, the shape of this model still shows confluence with the zero-crossing points of the classical and relativistic Compton-$y$ distortions, actually increasing the magnitude of covariances between $y$, $kT_\text{eSZ}$, the dust and CIB foreground parameters, and $A_\text{\cii{}}$ relative to the fiducial model, as~\cref{fig:cov_pad18} shows. The result is increased errors for these parameters as shown in~\autoref{tab:pad18}, with the main difference being that the relativistic corrections become more difficult to discern. Nonetheless, both the primary classical Compton-$y$ distortion and the \cii{} line emission remain detectable at high significance, so the qualitative opportunities highlighted with our fiducial models remain unchanged.}

Some of these conclusions will change also with incorporation of more LIM experiments covering more redshifts as well as \cii{} and higher-$J$ CO line emission, versus our particular CO LIM concept mostly covering $z\sim3$ and $z\sim7$ in specific low-$J$ CO lines. Other observables tracing the large-scale structure underlying the CO/\cii{} line emission will add even further information---see, e.g., the analysis of~\cite{Anderson22} considering FIRAS cross-correlation with galaxy surveys at $z\sim0.5$ and forecasting \PIXIE cross-correlations. \changeJfirstround{Our study thus only highlights some of the possible synergistic gains of a CMB spectrometer with LIM experiments, motivating more detailed studies for the future.}

\section{Conclusions}
\label{sec:conclus}

We now have answers to the questions posed towards the end of the~\href{sec:intro}{Introduction}:
\begin{itemize}
    \item \emph{How can spectral distortion surveys constrain the modelling of cosmological CO and \cii{} emission?} Given their broad frequency access, future concepts like \PIXIE and \SPIXIE can potentially achieve strong constraints on our simple model for CO and \cii{}, with \PIXIE (\SPIXIE) mission sensitivities being sufficient in principle for $\sim10\%$ ($\sim1\%$) \changeJ{level} constraints on molecular gas abundance and kinetic temperature evolution. Depending on the brightness of CO emission at high redshift, these constraints improve further with snapshots of CO emission at $z\sim3$ and $z\sim7$ provided by CO LIM experiments.
    \item \emph{How does the modelling of extragalactic CO and \cii{} in turn affect constraints on $y$- and $\mu$-type distortions?} In our forecasts for \PIXIE, introducing our fiducial model for CO and \cii{} increases the error on $y$ by around 10\% and the error on $\mu$ by a more considerable 50\%. Folding in detections from LIM observations can recover much of this increased uncertainty, \changeJfirstround{demonstrating the overall robustness of future CMB spectrometers with respect to more complex line emission modeling}. The extent of increase in error and the amount of information introduced by LIM data are both somewhat model dependent, but deviations in the CO/\cii{} global signal can clearly mimic $y$- and $\mu$-type distortions in combination with other foregrounds.
\end{itemize}

We note that there is considerable room to further develop consideration of foregrounds. In particular, as~\cref{fig:pathology} shows, it is the interaction of CO and \cii{} with other foreground components that amplifies their impact on spectral distortion errors. In particular, our models for Galactic and extragalactic dust are extremely simple and are far from reflecting the physical degrees of freedom in both the composition of Galactic dust and the distribution of extragalactic dust temperatures~\citep[cf., e.g.,][]{ZelkoFink21}. \changeJ{In addition, spatial variations of the spectral parameters across the sky will lead to new spectral shapes (i.e., foreground distortions) that can be captured using a moment-expansion method \citep{Chluba2017foregrounds}. The effect of these varying parameters will enter at second order in the average SEDs, and can be taken into account using extended ILC methods \citep{Rotti2021, Remazeilles2022, Rotti2022}. The same methods were successfully applied to $B$-mode forecasts \citep{Remazeilles2021, Azzoni2021, Carones2024}, demonstrating their potential, although more work is still needed in the context of CMB distortion science.}

However, we envision further consideration of different components can actually ameliorate their effect on the science of future microwave spectrometers. \changeJfirstround{As~\cref{fig:derivatives} indicates, the various LIM parameters can have similar overall effects on the associated monopole spectral template, which suggests a reduction of the dimensionality (e.g., by studying principal components) could further improve achievable precision in distortion parameters. In addition, $\omega_2^{\rm eSZ}$ (the second moment of the average relativistic SZ effect) contributes at a very small level to the expected distortions and \changeJ{can} thus be kept fixed or be predicted within a theoretical halo model approach \citep[e.g., {\tt class\_sz};][]{CLASS_SZ2023}, eliminating the need to marginalize over it. From the experimental point of view, a more careful optimization of the FTS, e.g., to improve the frequency sampling and sensitivity at high frequencies, would also enhance our ability to discriminate foregrounds. The use of spatial information from the CMB spectrometer itself could also provide additional gains \citep{Rotti2021}, consideration of which we omitted here. Our analysis should therefore be mainly taken as a first demonstration of the potential that a future CMB spectrometer holds in terms of LIM and distortion science, motivating numerous additional explorations.}

However, we expect certain aspects of our CO/\cii{} models to be generic. Given our knowledge of cosmic star-formation history and the low-$J$ CO lines, the CO global signal is overwhelmingly likely to peak somewhere between 100 GHz and 200 GHz, while the \cii{} global signal is likely to peak somewhere between 500 GHz and 1 THz and fall sharply beyond 300 GHz. These features should be sufficient for CO and \cii{} foregrounds to mimic relic spectral distortions as in~\cref{fig:pathology}, making their modelling of prime importance for science with CMB spectral distortions. This may result in bias of recovered $y$ and $\mu$ parameters even if the effect on uncertainties is small, which is beyond the scope of this work and its Fisher forecasts, \changeJfirstround{but should certainly be studied more carefully}.

Nonetheless, the CO/\cii{} global signal also presents a key opportunity for extragalactic astrophysics. \autoref{fig:breakdown} shows that (with the possible exception of $y$-type distortions) none of the other foregrounds we have considered quite has the spectral structure of the CO/\cii{} spectrum, with its non-monotonic behaviour between 100 GHz and 1 THz as the spectrum changes from CO-dominated to \cii{}-dominated, and its abrupt cutoff at the rest-frame emission frequency of the \cii{} line---all expected to be generic features. As such, a targeted search for the CO/\cii{} global signal in \PIXIE data is not out of the question, although it merits further study on the effects of systematics and foregrounds.

In both contexts, a key requirement for future work will be the ability to predict the CO/\cii{} global signal at speed. In this work, we used explicit non-LTE radiative transfer calculations to evaluate $L'_J(M_h)$. This is relatively fast if only a single evaluation is needed, but quickly becomes a computational bottleneck in contexts like Markov chain Monte Carlo inferences of parameters. As such, the original work of~\cite{COMAPTDX} used a surrogate model in the form of a Gaussian process emulator. One may also consider manual reduction of semi-analytic model outputs into empirical scaling relations~\citep{Yang22} or Gaussian mixture models of galaxy properties~\citep{Zhang23} for use with halo models, or even semi-numerical excursion set approaches that bypass resolving dark matter halos~\citep{LIMFAST1,LIMFAST2}. We expect these tools to develop naturally as the needs of the LIM community evolve, but a view towards unified modelling of correlated LIM observables across all redshifts would also strongly benefit the planning and optimisation of future \PIXIE-like or other spectrometric observations at microwave frequencies.

Finally, we note the strong complementarity of such a search in absolute spectrophotometric data with the ongoing development of CO and \cii{} LIM surveys. LIM surveys, with their finer angular and spectral resolution, will be able to capture the line-intensity anisotropies corresponding to the large-scale structure in specific survey volumes. The finer resolution provides additional spatial information that will aid separation of certain foregrounds, especially Galactic emission. However, analyses of LIM observations are likely to discard large longitudinal modes tracing the evolution of the background average, thus being chiefly sensitive to the scale of fluctuations rather than the cosmic average line intensity. Absolute spectrometers can potentially directly probe the latter with a sufficient understanding of the other foregrounds involved, given certain unique and generic aspects expected of the CO/\cii{} global signal. Joint analyses and direct cross-correlations in overlapping survey areas can only further help both types of surveys, introducing additional spatial information from LIM data and absolute calibration from spectrometer data. We hope future work will thoroughly explore these possibilities, including potential optimisation of the design and analysis both kinds of experiments to exploit such synergies, in the interest of both the CMB spectral distortion and LIM communities.

\vspace{3mm}
\section*{Acknowledgements}
\vspace{-2mm}
The authors thank Jos\'{e} Luis Bernal, Ely Kovetz, and other organisers of the `Present and Future of Line-Intensity Mapping' workshop at the Max Planck Institute for Astrophysics, for activating helpful discussions that ultimately encouraged this work.

DTC is supported by a CITA/Dunlap Institute postdoctoral fellowship. The Dunlap Institute is funded through an endowment established by the David Dunlap family and the University of Toronto. The University of Toronto operates on the traditional land of the Huron-Wendat, the Seneca, and most recently, the Mississaugas of the Credit River; DTC is grateful to have the opportunity to work on this land. DTC also acknowledges support through the Vincent and Beatrice Tremaine Postdoctoral Fellowship at CITA.
\changeJfirstround{This project also received funding from the European Research Council (ERC) under the European Union’s Horizon 2020 research and innovation program (grant agreement No 725456; CMBSPEC).
JC was furthermore supported by the Royal Society as a Royal Society University Research Fellow at the University of Manchester, UK.}
PCB is supported by the James Arthur Postdoctoral Fellowship.

\addedtwo{The authors also acknowledge and thank the service of an anonymous referee, whose attentive and constructive feedback improved this article.

}This work made use of \texttt{astropy},\footnote{http://www.astropy.org} a community-developed core Python package and an ecosystem of tools and resources for astronomy \citep{astropy:2013_, astropy:2018_, astropy:2022fix}. This research also made use of NASA's Astrophysics Data System Bibliographic Services.

\appendix





\bibliography{biblio,bibliold,bibliorig,biblico,nph-bib_query,Lit} 



\end{document}
\end{document}